\documentclass{article}

\usepackage{PRIMEarxiv}

\usepackage[utf8]{inputenc} 
\usepackage[T1]{fontenc}    
\usepackage{hyperref}       
\usepackage{url}            
\usepackage{booktabs}       
\usepackage{amsfonts}       
\usepackage{nicefrac}       
\usepackage{microtype}      
\usepackage{lipsum}
\usepackage{fancyhdr}       
\usepackage{graphicx}       
\graphicspath{{imgs/}}     
\usepackage{authblk}
\usepackage{amsmath}
\usepackage{optidef}
\usepackage{enumitem}
\usepackage{caption}
\usepackage{subcaption}
\usepackage{url}
\usepackage{mdframed}

\begingroup
\catcode`\_=\active
\gdef\mathurl#1{\text{\url{#1}}}
\endgroup

\let\oldurl\url
\renewcommand{\url}[1]{\text{\oldurl{#1}}}

\newcommand{\Graph}{{\mathcal{G}}}
\newcommand{\edges}{{\mathcal{E}}}
\newcommand{\vertices}{{\mathcal{V}}}

\title{Hybrid Quantum-Classical Branch-and-Price Method for the Vertex Coloring Problem
}

\author[1,2]{Chiara Vercellino}
\author[3]{M. Yassine Naghmouchi}
\author[3]{Wesley Coelho}
\author[1,2]{Giacomo Vitali}
\author[1]{Alberto Scionti}
\author[1]{Paolo Viviani}
\author[1]{Olivier Terzo}
\author[2]{Bartolomeo Montrucchio}

\affil[1]{Fondazione LINKS, Torino, IT \\ \texttt{firstname.lastname@linksfoundation.com}}
\affil[2]{DAUIN, Politecnico di Torino, Torino, IT \\ \texttt{firstname.lastname@polito.it}}
\affil[3]{PASQAL, Paris, FR \\ \texttt{firstname.lastname@pasqal.com}}

\begin{document}

\maketitle

\begin{mdframed}[linewidth=0.5pt]
This work has been submitted to the IEEE for possible publication. 
Copyright may be transferred without notice, after which this version 
may no longer be accessible.
\end{mdframed}
\vspace{1cm}

\begin{abstract}
This paper introduces Quantum Classical Branch-and-Price (QCBP), a hybrid quantum-classical algorithm for the Vertex Coloring problem on neutral-atom Quantum Processing Units (QPUs). QCBP embeds quantum computation within the classical Branch-and-Price (BP) framework to address three bottlenecks in classical BP algorithms: the computational cost of Pricing Subproblems (PSPs), branching efficiency, and the quality of primal heuristics. It uses quantum-assisted Column Generation (CG) based on Quantum Adiabatic Algorithms (QAA) to sample high-quality maximum-weight independent sets (MWIS), reducing the need to repeatedly solve NP-hard PSPs. The adapted branching strategy leverages quantum-generated independent sets to explore fewer nodes, tighten lower bounds, and converge faster. A classical primal heuristic rapidly builds feasible solutions from quantum-generated sets, avoiding unnecessary quantum calls or additional Integer Linear Programming (ILP) solves. Compared with our prior Hybrid Column Generation (HCG) and Branch-and-Bound through maximal Independent Set (BBQ-mIS), QCBP improves both quantum-resource utilization and solution quality. Extensive experiments show QCBP significantly outperforms HCG and BBQ-mIS, reaching optimality on $\approx 98\%$ of benchmark instances. Preliminary validation on real neutral-atom hardware indicates robustness to quantum noise and hardware constraints, supporting practical applicability and scalability to larger graph instances. QCBP emerges as a viable hybrid method for combinatorial optimization with promising scalability on near-term quantum hardware.
\end{abstract}


\section{Introduction}
\label{sec:introduction}
Quantum computing offers new potential for tackling combinatorial optimization problems that remain intractable for classical algorithms, especially those involving exponential search spaces. Among the various physical platforms, neutral atom quantum machines stand out for their scalability and natural ability to enforce hard constraints~\cite{henriet2020quantum}. These devices encode qubits using individual atoms held in place by optical tweezers, with quantum operations performed via precisely controlled laser pulses. The atomic states used typically correspond to a ground level and a highly excited Rydberg state. A central mechanism in this platform is the Rydberg blockade~\cite{ciampini2015ultracold}, which prevents simultaneous excitation of nearby atoms—enforcing a form of mutual exclusivity that maps directly to graph constraints. This hardware behavior makes neutral atom QPUs (Quantum Processing Units) particularly well-suited for solving the Maximum Independent Set (MIS) problem, which consists in selecting the largest possible subset of vertices in a graph such that no two are adjacent. When the input graph can be embedded as a unit-disk (UD) graph, the Rydberg blockade naturally enforces the independence condition, allowing direct mapping from problem constraints to physical interactions~\cite{dalyac2024graph, clark1990unit}. In contrast, other combinatorial problems—such as Vertex Coloring (VC) —cannot be enforced directly by the hardware. These problems require reformulation into an equivalent energy minimization form, such as Quadratic Unconstrained Binary Optimization (QUBO)~\cite{pichler2018quantum} model, which the quantum system can explore through variational or adiabatic evolution techniques. Solving such problems on neutral atom QPUs involves two main stages: (a) register design, where the problem graph/QUBO is mapped to a spatial layout of atoms respecting device constraints, and (b) pulse shaping, where the laser control parameters are optimized to guide the system toward low-energy configurations. Together, these capabilities are foundational for addressing even more complex tasks, such as Mixed-Integer Linear Programming (MILP)~\cite{naghmouchi2024mixed} and Integer Linear Programming (ILP) problems such as VC~\cite{da2023quantum, vercellino2023bbq}, through hybrid quantum-classical methods.

Vertex Coloring, an NP-hard combinatorial optimization problem, assigns colors to graph vertices such that adjacent vertices receive different colors, with the goal of minimizing the total number of colors used. It has diverse applications like scheduling, frequency assignment, and compiler optimization~\cite{demange2015some, ahmed2012applications, chiarandini2002application, lewandowski1994practical, formanowicz2012survey, chaitin1982register, leighton1979graph, hale1980frequency}. To handle VC’s complexity, both heuristics and exact algorithms have been developed. A particularly effective exact method is Branch-and-Price (BP), which becomes natural when the problem is formulated as a set partitioning ILP problem~\cite{hansen2009set}. In this formulation, each binary variable represents a feasible color, which is, an independent set (IS) of vertices that can receive the same color. However, the number of possible ISs in a graph grows exponentially with its size, making it impractical to enumerate all variables in advance. This is where Column Generation (CG) is used. Rather than considering all variables at once, CG starts with a small subset and solves a relaxed version of the problem, namely the Restricted Master Problem (RMP), where the integrity constraints of the binary variables is relaxed. It then iteratively adds new variables, called columns, that are most likely to improve the solution. These columns are identified by solving the so-called Pricing Subproblem (PSP), which involves finding a weighted MIS guided by the current dual solution. This step is itself NP-hard, but it avoids dealing with all possible MISs upfront. 

Once CG produces a good fractional solution, stemming from the fact that the RMP has only continuous variables, Branch-and-Bound (BB) is used to search the solution space for the optimal integer solutions. In parallel, a primal heuristic is called at the root and within BB nodes to build an integer coloring from the current column pool—completing uncovered vertices via on-the-fly pricing when needed. The heuristic supplies incumbents (tight upper bounds), enables early variable fixing, prunes the search tree, and often stabilizes CG by guiding column selection. BB systematically explores the solution space by branching on decisions such as which vertices should or should not share the same color. Various branching strategies exist for the VC, including branching on individual variables~\cite{malaguti2011exact}, edges~\cite{mehrotra1996column}, vertices~\cite{mendez2006branch}, or subsets of vertices~\cite{morrison2014wide}.

An efficiently implemented BP algorithm, can significantly reduce the search space and deliver high-quality solutions for VC~\cite{archetti2014branch, feillet2010tutorial}. However, BP algorithms may face significant challenges due to the inherent complexity of their components. The PSP, during CG, is computationally demanding, often involving NP-hard problems, such as MIS problems, that must be solved repeatedly. Additionally, the exponential growth of the solution tree requires carefully designed branching rules to efficiently explore the search space. Finally, the relative gap between the lower bound (LB) provided by CG and the best known integer solution highlights the need for primal heuristics, i.e., strategies that provide high-quality feasible integer solutions, during the search process. These solutions help refine upper bounds (UBs), guide the search, and accelerate convergence. The algorithm iteratively refines both bounds and terminates when the relative gap is reduced to zero or an acceptable threshold, ensuring optimality or near-optimality. In this paper, we explore the potential of quantum computing with neutral atom QPUs to enhance these aspects.

Recent efforts have explored hybrid quantum-classical approaches to tackle the VC problem, illustrating the potential of quantum routines to address the computational challenges discussed earlier. Two notable methods are our previous work, BBQ-mIS~\cite{vercellino2023bbq} and Hybrid Column Generation (HCG)~\cite{da2023quantum}. The BBQ-mIS method integrates quantum routines within a BB framework. At each node of the BB tree, quantum computations are used to generate maximal independent sets (mIS), which are then combined to build feasible colorings. The method uses classical pruning and bounding techniques to guide the resolution tree exploration. While effective for UD graphs, BBQ-mIS faces scalability issues on larger general graphs due to the amount of BB nodes that must be processed. On the other hand, the HCG method enhances CG by solving the PSP using quantum samplers. New independent sets are identified via quantum routines and injected into the RMP, which is solved classically. Although this reduces the total number of quantum iterations, the quality of the generated columns often degrades on general graphs, leading to suboptimal colorings. These limitations—BBQ-mIS suffering from an expensive BB exploration, and HCG from degraded solution quality—highlight the need for a unified hybrid approach that combines the strengths of both strategies while mitigating their weaknesses, particularly for general graph instances.

In this paper, we propose a quantum-classical BP approach (QCBP) to address these limitations and solve the VC problem more efficiently. Our contributions are as follows: (1) we formulate VC as a set partitioning problem and develop a BP algorithm that uses quantum-assisted column generation inspired by HCG~\cite{da2023quantum}; (2) we adapt the branching scheme from BBQ-mIS~\cite{vercellino2023bbq} to operate on dynamically generated columns; (3) we introduce a classical greedy primal heuristic that constructs feasible colorings by combining independent sets generated during CG, without additional QPU or ILP calls; (4) we optimize quantum parameters such as the number of shots and choice of quantum algorithm to ensure efficient integration with the classical solver and execution on real QPUs; and (5) we demonstrate experimentally that QCBP outperforms BBQ-mIS and HCG in terms of both solution quality and quantum resource usage across a diverse set of graph instances.

The structure of the paper is as follows: Section \ref{sec:related_work} reviews related work on BP methods, covering both classical and hybrid quantum-classical algorithms. Section~\ref{section:background} introduces the fundamental concepts of quantum optimization algorithms relevant to our approach and outlines the general framework of the BP method. Section~\ref{sec:problem_statement} presents the mathematical formulations associated with the VC problem, along with the necessary definitions. Section \ref{sec:qcbp} presents the proposed QCBP method. Section \ref{sec:results} compares the performance and efficiency of our methods both through classical emulation and using \textit{Orion Alpha}'s QPU \cite{pasqal_fresnel_job} (\textit{Pasqal} neutral atoms quantum device). Finally, Section \ref{sec:conclusion} summarizes the findings and outlines future research directions.

For a comprehensive list of acronyms used throughout this paper, the reader is refereed to Table~\ref{tab:acronyms}.

\begin{table}[ht]
\centering
\caption{List of Acronyms}
\begin{tabular}{ll}
\toprule
\textbf{Acronyme} & \textbf{Signification} \\
\midrule
BP         & Branch-and-Price \\
BB         & Branch-and-Bound \\
CG         & Column Generation \\
RMP        & Restricted Master Problem \\
PSP        & Pricing Subproblem \\
VC         & Vertex Coloring \\
VCPF       & Vertex Coloring set Partitioning Formulation \\
MIS        & Maximum Independent Set \\
mIS        & maximal Independent Set \\
IS         & Independent Set \\
QUBO       & Quadratic Unconstrained Binary Optimization \\
ILP        & Integer Linear Programming \\
LB         & lower bound\\
MILP       & Mixed-Integer Linear Programming \\
QPU        & Quantum Processing Unit \\
QCBP       & Quantum-Classical Branch-and-Price \\
BBQ-mIS    & Branch-and-Bound with Quantum-enhanced mIS generation \\
HCG        & Hybrid Column Generation \\
QAA        & Quantum Adiabatic Algorithm \\
QAOA       & Quantum Approximate Optimization Algorithm \\
VQE        & Variational Quantum Eigensolver \\
UB         & upper bound\\
UD         & unit-disk (graph) \\
NISQ       & Noisy Intermediate-Scale Quantum (devices) \\
DEN        & Distance Encoder Network \\
ELF        & Embedding Loss Function \\
\bottomrule
\end{tabular}
\label{tab:acronyms}
\end{table}

\section{Related work}
\label{sec:related_work}
Hybrid classical-quantum methods have recently emerged as promising strategies for addressing the computational complexity inherent to combinatorial optimization problems. Traditionally, BB algorithms systematically explore the solution space by branching on decision variables and applying bounding techniques to prune suboptimal regions of the solution space. This guarantees convergence towards optimal solutions while managing computational complexity. Recent developments extend this classical approach by incorporating quantum computation into hybrid classical-quantum BB frameworks. Sanavio et al.~\cite{sanavio2024hybrid} introduced such a hybrid approach, employing quantum algorithms to efficiently resolve challenging subproblems like the knapsack and traveling salesman problems within the BB process. Similarly, our prior work, BBQ-mIS~\cite{vercellino2023bbq}, demonstrated the integration of quantum routines within a BB framework specifically designed for VC on unit-disk graphs, leveraging quantum algorithms to identify maximal independent sets at each branching node. In that work, we used the Quantum Approximate Optimization Algorithm (QAOA)~\cite{farhi2014quantum} to solve MIS problems on neutral atom quantum processors and post-selected the solutions corresponding to mISs. QAOA combines quantum and classical optimization steps, with parameters tuned using Bayesian optimization~\cite{skopt}. On analog hardware, the algorithm is implemented via a time-varying Hamiltonian, leveraging the Rydberg blockade effect~\cite{ciampini2015ultracold} to encode the problem constraints. The main limitation of BBQ-mIS is its high demand for quantum resources, due to the significant number of QPU calls involved in the QAOA phase (an iterative algorithm) and the lower bounds, which do not guarantee fast convergence

In a broader sense, classical BB algorithms inherently struggle when addressing optimization problems characterized by exponentially expanding solution spaces. To mitigate these limitations, decomposition techniques such as Benders Decomposition~\cite{bnnobrs1962partitioning, zhao2022hybrid, gao2022hybrid, chang2020hybrid} and CG~\cite{mehrotra1996column} have been extensively explored. These methods strategically partition complex optimization problems into smaller, more manageable subproblems, thus significantly enhancing computational tractability.

For optimization tasks characterized by exponential growth in their variable spaces, typical of VC problems, CG has proven particularly effective. CG operates by iteratively solving a RMP; the linear relaxation of the original problem, while introducing new variables through solving computationally demanding PSPs. Hybrid quantum-classical CG methods integrate quantum algorithms into this step, exploiting quantum efficiency to accelerate PSP solutions. Ossorio-Castillo and Peña-Brage~\cite{ossorio2022optimization}, for instance, proposed a hybrid CG method applied to refinery scheduling, employing quantum annealing to solve PSPs formulated as QUBO problems. Similarly, the HCG method by da Silva et al.~\cite{da2023quantum} utilized quantum routines effectively within CG to enhance performance in VC problems. However, these quantum-enhanced CG methods often face scalability issues stemming from hardware limitations and do not inherently ensure integer-optimal solutions.

To address the shortcomings of both pure CG and classical BB methods, BP techniques combine the strengths of BB and CG.
BP algorithms have historically demonstrated notable efficiency and applicability across diverse sectors, including transportation, logistics, energy, and production planning~\cite{yildiz2016branch, dell2006branch, bard2010branch, anjos2017unit, gronhaug2010branch, morvaj2016optimization, lopes2007branch}. Furthermore, BP approaches are effective in tackling complex cutting and packing problems, such as telecommunications network design~\cite{xu2022branch, naghmouchi2023optimal}, and service and healthcare optimization problems~\cite{belien2008branch}.

With the advent of practical quantum computing technologies, hybrid quantum-classical BP methods are emerging as promising strategies to address even larger and more challenging optimization problems. Quantum algorithms, including the QAOA~\cite{choi2019tutorial} and Variational Quantum Eigensolver (VQE)~\cite{tilly2022variational}, offer significant promise in solving computationally intensive subproblems within hybrid optimization frameworks. Recent examples of such approaches include the work by Svensson et al.~\cite{svensson2023hybrid}, who explored a hybrid BP algorithm applying QAOA within the RMP context, and Tran et al.~\cite{tran2016hybrid}, who employed quantum annealing to tackle complex scheduling subproblems. Distinct from prior works, our approach uniquely integrates neutral atoms quantum computing within the PSP rather than the RMP.

Motivated by these insights, this paper introduces a hybrid quantum-classical BP algorithm, designed explicitly to overcome limitations encountered by current hybrid CG and BB methods. By integrating quantum-enhanced PSP solving capabilities within the strategic exploration framework of classical BP, the proposed approach addresses general VC instances with exponential solution spaces.

\section{Background}
\label{section:background}
In this section, we provide the essential background on neutral‐atom QPUs and the classical BP framework to set the stage for our hybrid algorithm.

\subsection{Neutral atoms QPUs for solving optimization problems}

Neutral atom QPUs leverage individual atoms as qubits, controlled through laser-based manipulation. Each atom encodes qubit states, typically the ground state ($|g\rangle$) and the Rydberg state ($|r\rangle$), which are manipulated using precisely tuned lasers. The key mechanism underlying these QPUs is the Rydberg blockade, which ensures that two atoms within a certain proximity cannot be simultaneously excited to the Rydberg state. This mechanism naturally enforces constraints in optimization problems.
The system's dynamics are governed by a Hamiltonian, $H(t)$, parameterized by the Rabi frequency $\Omega(t)$, the laser detuning $\delta(t)$, and atomic interaction terms $U_{ij}$. By shaping these parameters over time, the QPU processes quantum information and evolves toward optimal solutions.

Let $n$ be the number of atoms in the register, $\Omega(t)$ the Rabi frequency, $\delta(t)$ the laser detuning, $r_{ij}$ the Euclidean distance between atoms $i$ and $j$, and $\sigma_i^x$ the Pauli-$X$ operator acting on atom $i$. We define $n_i$ as the occupation number of atom $i$, with $n_i = 1$ if the atom is in the excited state and $n_i = 0$ otherwise. The time-dependent Ising Hamiltonian governing the system is given by:

\begin{equation}
H(t) = \Omega(t) \sum_{i=1}^{n} \sigma_i^{x} 
       - \delta(t) \sum_{i=1}^{n} n_i 
       + \sum_{1 \leq i < j \leq n} \frac{C_6}{r_{ij}^6} \, n_i n_j ,
\label{eq:rydberg_hamiltonian}
\end{equation}

where the term $\frac{C_6}{r_{ij}^6} n_i n_j$ represents the interaction energy between atoms $i$ and $j$, effectively encoding the coupling strengths $U_{ij}$.

The process of solving optimization problems using neutral atom QPUs involves two key phases: problem encoding via register desing and problem solving via pulse shaping.

\subsubsection{Problem Encoding via Register Design}

In the encoding phase, the problem is mapped onto the QPU register through a process called register design. This involves spatially arranging the atoms such that their interactions $U_{ij}$ represent the off-diagonal terms of a QUBO instance or the connectivity of the graph defining the VC problem. The embedding quality is critical, as it directly impacts the accuracy and feasibility of the solution. 

The embedding process often relies on heuristics, as finding optimal atomic arrangements is computationally challenging, especially for large graphs~\cite{naghmouchi2023optimal, vercellino2023neural}. Even for specific graph structures, such as unit-disk graphs, exact embeddings might be challenging. As a result, approximate embeddings are commonly used, which may lead to suboptimal interactions.
In this work, we employ a neural network–based heuristic, the Distance Encoder Network (DEN) model~\cite{vercellino2023neural}, to obtain UD representations of input graphs that are compatible with our target QPU. It is important to note that not all graphs admit a UD representation, as this imposes geometric constraints on node placement and connectivity. In cases where an exact UD representation is not feasible, we resort to non-UD embeddings. These satisfy QPU placement requirements (detailed in \ref{sec:register_embedding}), however, the resulting interaction patterns may deviate from those of the original problem graph.

\subsubsection{Problem Solving via Pulse Shaping}

After encoding the problem, the QPU evolves the quantum state to find a solution. This evolution is guided by a carefully designed sequence of laser pulses, a process known as pulse shaping. Depending on the encoding accuracy, two approaches are employed.

For exact embeddings, adiabatic pulses are designed in accordance with the adiabatic theorem, ensuring a smooth and continuous evolution from an easily prepared initial state to the ground state of the target Hamiltonian. This is achieved by gradually varying key parameters such as the Rabi frequency $\Omega(t)$ and the laser detuning $\delta(t)$ (see \eqref{eq:rydberg_hamiltonian}), thereby maintaining the system in its instantaneous ground state throughout the evolution.
Importantly, suitable values for $\Omega(t)$ and $\delta(t)$ can be derived from the atomic positions defined by the UD representation of the graph~\cite{da2022efficient, vercellino2022neural}, as this spatial layout determines which atoms fall within the Rydberg interaction radius and thus which interactions are relevant during evolution. This adiabatic approach underpins the Quantum Adiabatic Algorithm (QAA)~\cite{aqmis}, which has proven effective in solving combinatorial optimization problems such as the MIS, particularly on UD graphs. In this context, QAA gradually transforms the system from the ground state of an easily prepared initial Hamiltonian $H_0$ to the ground state of the problem Hamiltonian $H_C$, whose parameters are determined by the problem instance. In the MIS problem, the Rydberg blockade mechanism naturally enforces the independent set constraints, enabling the quantum system to explore feasible configurations and converge to high-fidelity solutions.


When exact embeddings are infeasible, variational quantum algorithms can be employed. In this approach, non-adiabatic pulses are designed to approximate the desired Hamiltonian dynamics. Variational methods, such as the QAOA~\cite{farhi2014quantum}, are used to optimize pulse parameters iteratively. These methods are particularly useful for noisy or imperfect quantum systems, where achieving exact solutions may not be possible. QAOA alternates between the cost Hamiltonian $H_C$ and a mixer Hamiltonian $H_M$ to explore the solution space. It optimizes the parameters of these Hamiltonians through classical feedback, enabling efficient problem-solving even with approximate embeddings. By carefully selecting the number of QAOA layers and pulse durations, the algorithm balances solution quality with computational feasibility.

While QAOA provides broader flexibility for general graphs, we choose to rely on the QAA for our QCBP method, as it aligns more closely with the strengths of neutral atom QPUs. These platforms offer inherent scalability and noise resilience due to their analog nature, reducing error propagation even under imperfect embeddings. Although QAA is less tolerant to embedding deviations than QAOA, its lower overhead, in terms of quantum resources, makes it attractive for near-term devices. Whereas exact representations are ideal, QAA remains effective in approximate settings, as we leverage it not as a strict quantum solver but as a quantum sampler. Even suboptimal solutions generated by QAA can serve as valuable inputs to our QCBP method, supporting the broader optimization process within our hybrid framework.







\subsection{Classical Branch-and-Price Algorithm}

The BP algorithm~\cite{feillet2010tutorial, savelsbergh1997branch, vanderbeck2011branching, sadykov2013bin} integrates CG within a BB framework, efficiently solving large-scale ILPs. It leverages CG for dynamic variable generation, a primal heuristic to enforce integrality, and a BB scheme to guide the search in the solution space. Formally, we consider an ILP defined as:
\begin{align}
    P: \quad \min & \quad c^\top x \\
    \text{s.t.} & \quad Ax \geq b, \\
    & \quad x \in \{0,1\}^n,
\end{align}
where $ x \in \{0,1\}^n $ represents binary decision variables, $ A \in \mathbb{R}^{m \times n} $ is the constraint matrix, $ b \in \mathbb{R}^{m} $ the constraint vector, and $ c \in \mathbb{R}^{n} $ the cost vector.

Solving $P$ directly is often impractical for large-scale instances due to the exponential number of potential variables. Thus, BP begins by solving the linear relaxation $P_{LR}$, defined as:
\begin{align}
    P_{LR}: \quad \min & \quad c^\top x \\
    \text{s.t.} & \quad Ax \geq b, \\
    & \quad x \geq 0.
\end{align}

Due to the large size of $P_{LR}$, CG iteratively solves an RMP, a reduced version of $P_{LR}$ that includes only a subset of the variables (columns). At iteration $k$, the RMP is formulated as:
\begin{align}
    \text{RMP}^{(k)}: \quad \min & \quad (c^{(k)})^\top x^{(k)} \\
    \text{s.t.} & \quad A^{(k)} x^{(k)} \geq b, \\
    & \quad x^{(k)} \geq 0,
\end{align}
where $ A^{(k)} \in \mathbb{R}^{m \times n_k}$ is a submatrix of $ A $ corresponding to the current subset of variables, and $ c^{(k)} \in \mathbb{R}^{n_k}$ is the corresponding subset of the cost vector.

After solving the primal RMP$^{(k)}$, we retrieve the dual optimal solution $ y^{(k)} \in \mathbb{R}^{m}$, which is essential for the PSP. The PSP identifies new columns (variables) that potentially improve the current solution, by computing the reduced costs based on dual prices $y^{(k)}$. Formally, the reduced cost $\bar{c}_j$ of a candidate column $A_j$ (variable $x_j$) is given by:
\[
\bar{c}_j = c_j - A_j^\top y^{(k)}.
\]

The PSP seeks variables with negative reduced costs ($\bar{c}_j < 0$). If such variables are found, they are added to the RMP, and the CG procedure continues to the next iteration. Otherwise, optimality for the linear relaxation is reached, and CG terminates.

Since the solution obtained through CG may be fractional, a primal heuristic is employed to retrieve an integer solution.
As the feasible integer solution is obtained heuristically, a BB framework is needed to guide the search process towards optimality.

\begin{enumerate}
    \item \textbf{Branching Step:}  
    A fractional variable $x_j$ from the current optimal solution is selected for branching. Typically, branching splits the current problem into two subproblems by adding constraints such as $ x_j = 0 $ and $ x_j = 1 $. However, in practice, more sophisticated branching strategies, such as branching on constraints or applying problem-specific strategies, can be employed to accelerate convergence.

    \item \textbf{Bounding Step:}  
    Each generated subproblem’s linear relaxation is solved iteratively using CG, yielding lower bounds on the objective function. Nodes whose bounds exceed the current best integer solution (upper bound, UB) are pruned.
\end{enumerate}

The algorithm recursively applies these steps, constructing and exploring a tree of subproblems until all feasible solutions are identified or excluded. The BP procedure terminates when all nodes in the search tree have either been explored or pruned, and an integer solution with proven optimality has been obtained. With these foundations, in the following section, we formalize the VC problem and introduces the ILP formulations essential for our hybrid BP approach.

\section{Vertex Coloring: Problem Statement and ILP Formulations}
\label{sec:problem_statement}
This section defines the VC problem and details both the compact and set-partitioning ILP formulations that underpin our hybrid approach.
We consider the classical VC problem. Let $\mathcal{G}(\mathcal{V}, \mathcal{E})$ be an undirected graph, where $\mathcal{V}$ is the set of $n$ vertices and $\mathcal{E}$ the set of edges. The objective of the VC problem is to assign a color to each vertex such that no two adjacent vertices share the same color, using the minimum number of colors possible. This minimum number is called the chromatic number and is denoted by $\chi(\mathcal{G})$. In other words, a feasible coloring is a partition of the vertex set into independent sets (ISs), and $\chi(\mathcal{G})$ corresponds to the smallest number of such independent sets required to cover $\mathcal{V}$. An IS is a subset of vertices in which no two vertices are adjacent. Among them, a \textit{maximal} independent set is an IS that cannot be extended by including any other vertex from $\mathcal{V}$ without violating independence. A \textit{maximum} independent set is the largest set among the all possible mISs on the graph.

Several integer programming formulations have been proposed for the VC problem. A collection of such formulations can be found in~\cite{hansen2009set}. In the following, we present two prominent approaches: a compact formulation and a set partitioning formulation based on column generation.

\subsection{Compact ILP Formulation}

A standard ILP formulation for the VC problem introduces binary decision variables $x_{ic}$, which equal 1 if vertex $i \in \mathcal{V}$ is assigned color $c \in \mathcal{M}$, and $y_c$, which equals 1 if color $c$ is used in the solution. Here, $\mathcal{M}$ denotes a fixed set of $m$ candidate colors, where $m$ is an upper bound on $\chi(\mathcal{G})$ (for example, $m=n$). The ILP reads as follows:

\begin{mini!}|s|
{}{\sum_{c=1}^m y_c }{}{}
\addConstraint{\sum_{c=1}^m x_{ic} = 1}{\quad \forall i \in \mathcal{V}} \label{eq:compact:one_color}
\addConstraint{x_{ic} + x_{jc} \leq y_c}{\quad \forall (i,j) \in \mathcal{E},\ \forall c \in \mathcal{M}} \label{eq:compact:adjacent}
\addConstraint{x \in \{0,1\}^{n \times m}}{}
\addConstraint{y \in \{0,1\}^{m}}{}
\end{mini!}

Constraint \eqref{eq:compact:one_color} ensures that each vertex is assigned exactly one color, while constraint \eqref{eq:compact:adjacent} guarantees that adjacent vertices do not receive the same color. This compact formulation is widely used in benchmarking because of its simplicity and general applicability. However, it is known to have a weak linear relaxation, which often results in poor lower bounds and makes it computationally challenging for large or dense graphs. Furthermore, the simplest corresponding QUBO formulation, where the constraints in \eqref{eq:compact:one_color} are enforced via off-diagonal penalty terms and the minimization over the $y$ variables is performed iteratively, requires significantly more qubits, specifically $n \times m$, compared to the $n$ qubits needed in the PSP formulation. This limitation is particularly critical when operating on NISQ devices, which have a limited number of qubits. By optimizing qubit usage, we can therefore tackle more challenging problem instances effectively.

\subsection{Set Partitioning Formulation}

A powerful formulation of the VC problem can be based on covering the graph using ISs that partition the vertex set. Let $\mathcal{S}$ denote the set of all ISs of the graph $\mathcal{G}$. We introduce a binary variable $\lambda_S$ for each $S \in \mathcal{S}$, where $\lambda_S = 1$ if and only if the set $S$ is selected in the coloring.

The VC problem can then be formulated as the following ILP denoted as VCPF:

\begin{mini!}|s|
{}{\sum_{S \in \mathcal{S}} \lambda_S}{}{}\label{eq:cg:obj}
\addConstraint{\sum_{S \ni i} \lambda_S = 1}{\quad \forall i \in \mathcal{V}} \label{eq:cg:partition}
\addConstraint{\lambda_S \in \{0,1\}}{\quad \forall S \in \mathcal{S}} \label{eq:cg:binary}
\end{mini!}

Constraint \eqref{eq:cg:partition} ensures that each vertex is assigned to exactly one selected IS. Since no two adjacent vertices can belong to the same independent set, this enforces a valid coloring of the graph, where each set corresponds to a distinct color class. This is the formulation of problem $P$ for the VC problem.

Due to the exponential number of ISs in general, this formulation is typically addressed using a column generation approach. The master problem starts with a limited subset of $\mathcal{S'}$, and new columns are generated on-the-fly by solving a pricing problem. The PSP consists of identifying one or more ISs with negative reduced cost. 

The linear programming relaxation, equivalent to $P_{LR}$ formulation, is VCPF-R, and it is obtained by replacing the integrality constraints~\eqref{eq:cg:binary} with non-negativity constraints:

\begin{mini!}|s|
{}{\sum_{S \in \mathcal{S'}} \lambda_S}{}{}
\addConstraint{\sum_{S \ni i} \lambda_S = 1}{\quad \forall i \in \mathcal{V}} \label{eq:cg:relaxed_partition}
\addConstraint{\lambda_S \geq 0}{\quad \forall S \in \mathcal{S'}} \label{eq:cg:relaxed_nonneg}
\end{mini!}

When solving this linear program, the dual formulation naturally arises and introduces a \textit{dual} variable $\pi_i$ for each vertex $i \in \mathcal{V}$, leading to the following dual formulation:

\begin{maxi!}|s|
{}{\sum_{i \in \mathcal{V}} \pi_i}{}{}
\addConstraint{\sum_{i \in S} \pi_i \leq 1}{\quad \forall S \in \mathcal{S'}} \label{eq:cg:dual_constraint}
\addConstraint{\pi_i \in \mathbb{R}}{\quad \forall i \in \mathcal{V}} \label{eq:cg:duals_domain}
\end{maxi!}

The dual variables assign weights to each vertex, representing the marginal cost or value of including that vertex. These dual variables are essential for guiding the column generation procedure. Since smaller reduced costs correspond to more valuable ISs for solving the RMP, the PSP for the VC problem is modeled as a Maximum Weight Independent Set (MWIS) problem, with vertex weights defined by the dual variables $\pi$ from the previous iteration.
In the next section, we present our hybrid quantum-classical BP algorithm built upon these mathematical models.

\section{QCBP algorithm for VC problems}\label{sec:qcbp}
The hybrid BP algorithm consists in integrating a hybrid CG procedure within a BB framework. 

\subsection{Hybrid Column Generation algorithm}\label{sec:hcg}

The aim of HCG is to find a high quality solution to VCPF-R with a restricted number of variables. At each iteration of CG, the procedure solves a linear program with a limited subset of variables associated with $\mathcal{S}' \subseteq \mathcal{S}$, which is the RMP. The initial RMP is typically initialized with a small set of feasible independent sets, e.g., singletons for each vertex. Once a solution of the RMP is found, the PSP is solved to identify the variables that can improve the objective function of VCPF-R, and which must be added to the RMP. The RMP updated with these new variables is then re-solved, and the procedure repeats until no variable improving the objective function can be identified by the pricing problem. In this case, the solution obtained from the last RMP is the best for VCPF-R.

For a given RMP associated with a subset of independent sets $\mathcal{S}' \subseteq \mathcal{S}$, solving the PSP amounts to identifying a violated constraint in the dual of VCPF-R. Let $\pi_i$ be the dual variable associated with the primal constraint ensuring that vertex $i \in \mathcal{V}$ is covered exactly once. We denote by $\pi^* = (\pi^*_i)_{i \in \mathcal{V}}$ the current values of the dual variables obtained from the RMP. From the dual formulation of VCPF-R, it follows that the PSP reduces to finding at least one independent set $S^* \in \mathcal{S} \setminus \mathcal{S}'$ such that
\begin{align}
\label{eq:pricing_violation}
\sum_{i \in S^*} \pi^*_i > 1.
\end{align}

Equivalently, the PSP consists of finding an independent set $S^*$ that maximizes the dual weight $\sum_{i \in S^*} \pi^*_i$. If the optimal value of this subproblem is greater than 1, then the corresponding column has negative reduced cost and is added to the master problem.

Therefore, the pricing subproblem reduces to solving the following:
\[
\max\left\{ \sum_{i \in S} \pi^*_i \ : \ S \subseteq \mathcal{V},\ S \text{ is an independent set} \right\}.
\]
To model it, we use the following ILP formulation:

\begin{mini!}|s|
{}{\sum_{i=1}^n \pi^*_i z_i \label{eq:objectif}}{}{}
\addConstraint{z_i + z_j \leq 1 \quad \forall (i,j) \in \mathcal{E} \label{eq:conflit}}{}
\addConstraint{z \in \{0,1\}^n \label{eq:binaire}}{}
\end{mini!}

With $z_i$ being the binary variable that takes the value 1 if the corresponding vertex is selected in the optimal solution, and $w_i = \pi^*_i$ the weight derived from the dual variables.

This is an MWIS problem, which is itself NP-hard. This motivates the use of quantum computing to solve it. To this end we adopt, with some implementation details adjustment, the approach proposed in~\cite{da2023quantum}. This framework benefits from a hybrid classical-quantum approach, where classical optimization techniques manage the master problem, while the neutral atom-based quantum solver addresses the combinatorial nature of the IS-oriented subproblems, which are naturally embedded into this type of quantum device. As highlighted in~\cite{da2023quantum}, the proposed algorithm remains a heuristic approach to solving the VC problem. However, a key advantage of using quantum methods for MWIS identification is the ability to generate multiple candidate solutions per iteration. Unlike exact ILP solvers that typically return a single optimal solution, quantum algorithms naturally support repeated sampling from a distribution over low-energy states. This enables the CG procedure to potentially identify several negative reduced-cost ISs in one iteration, accelerating convergence. Whereas it can provide high-quality solutions for a wide range of problem instances, embedding this method into a BP framework is necessary to seek optimality for any instance of the problem under consideration.

The hybrid methods presented in this paper rely on solving MWIS problems within the quantum component, as detailed in Section~\ref{sec:quantum_mis}. In most iterations, these pricing subproblems are addressed using QAA. While this enables efficient CG, it may fail to identify all negative reduced-cost columns, especially in later stages of the algorithm, due to the imperfect UD representation of the intended graph in the quantum register and the lack of optimality guarantees in QAA algorithms. To ensure the termination condition of the CG process is met correctly, we incorporate a safeguard: in the final iteration, if no improving columns are found via the quantum routine, the MWIS subproblem is solved classically and exactly using the \textit{GLPK} ILP solver~\cite{glpk}. This guarantees that no valid column has been missed and that the current solution is indeed optimal with respect to the relaxed master problem. Even though the use of an exact solver might seem at odds with the quantum approach, it is worth noting that the ILP solver is typically invoked at the end of the HCG procedure—precisely when the QPU and the QAA algorithm begin to struggle in finding reduced-cost ISs, which usually correspond to smaller subgraphs of $\Graph$ and are no longer a bottleneck for classical resources.

\begin{figure}[t]
\centering
\includegraphics[width=0.7\textwidth]{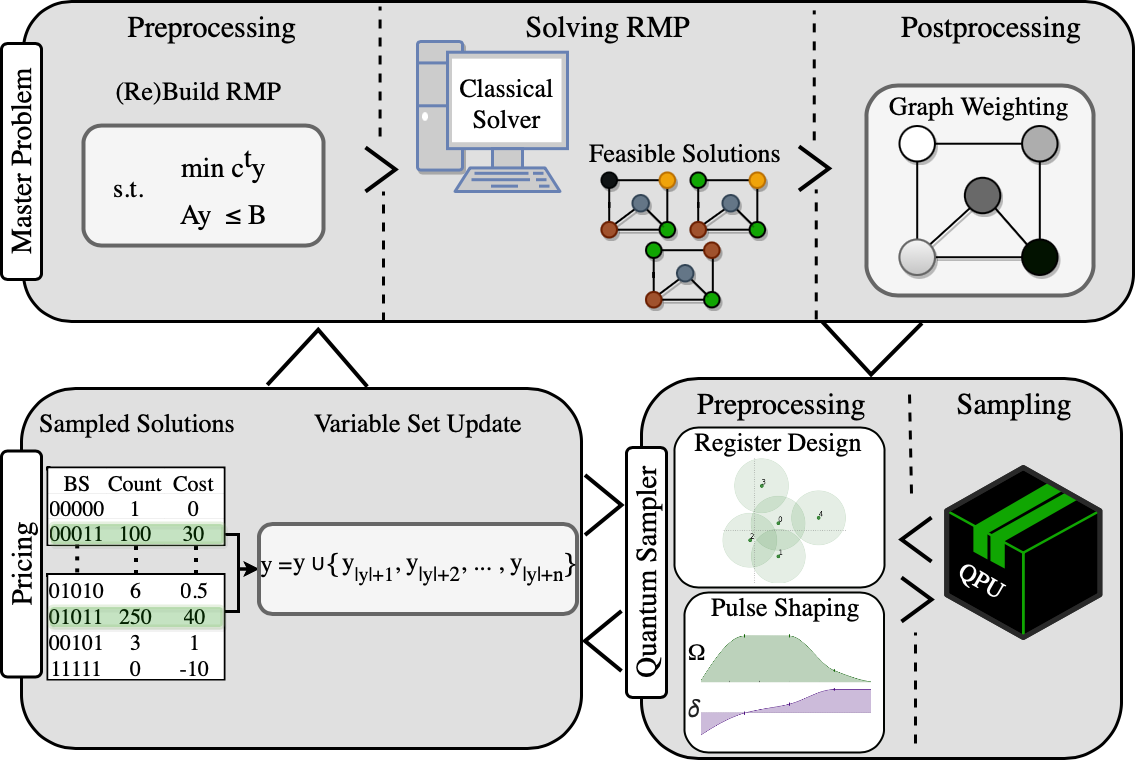}
\caption{Workflow of the hybrid classical-quantum column generation approach. First, a minimal sub-set $y$ of variables is generated in such a way that it ensures a feasible solution for the RMP (e.g., with only the singletons of the graph). The RMP is then solved by a classical solver.  The next steps are related to the pricing sub-problems, which are solved by considering the dual values from the solved RMP in order to find more variables that can potentially improve the current solution of the RMP. If such variables exist, then  RMP is updated with the new variables and is solved again. The search for new variables is done by a quantum sampler specifically tailored to consider different inputs related to each pricing iteration. These last steps are repeated until no column is generated by the PSP. Source~:~Adapted from~\cite{da2023quantum}.}\label{workflowintro}
\end{figure}

\subsection{Quantum routines for solving the PSP} \label{sec:quantum_mis}
Here, we present the quantum procedures for tackling the MWIS PSP within our hybrid BP framework. We detail the two key quantum routines, i.e., register design and pulse shaping, that describe how we embed the graph into a neutral-atom register and employ adiabatic pulse to efficiently sample high-weight independent sets.

\subsubsection{Graph embedding heuristic}\label{sec:register_embedding}

Register design is a fundamental step that we performed to effectively represent VC problem instances on a neutral atom QPU. Specifically, our method exploits the Rydberg blockade effect by strategically positioning atoms within the quantum register and carefully selecting the Rabi frequency $\Omega$ to represent the edges of the graph. This approach inherently corresponds to unit-disk graph representations. However, since not all graphs naturally admit UD representations, the embedding step itself constitutes a challenging optimization problem, known to be NP-hard in general~\cite{breu1998unit}. Given the non-convex nature of this embedding problem and its potential requirement for integer variables, exact methods quickly become computationally infeasible, even for relatively small graphs (e.g., graphs with around $n \approx 10$ vertices). Therefore, we turned to heuristic methodologies to efficiently generate feasible embeddings.

Following the neural network-based approach initially proposed in~\cite{vercellino2022neural,vercellino2023neural}, we developed and implemented a dedicated deep learning framework tailored for this embedding problem. Our neural network model, known as DEN, computes pairwise distances between atoms (qubits), while our specifically crafted loss function, called ELF (Embedding Loss Function), enforces embedding constraints during training. This combination enables the identification of atom arrangements within the quantum register that satisfy both the UD criteria and hardware-specific constraints. In particular, the ELF loss function integrates multiple constraints, such as:
\begin{itemize}
\item The UD condition, defining the maximum allowed distance between adjacent atoms (approximately $10~\mu m$), essential for respecting the coherence properties of the quantum hardware.
\item Minimum spacing requirements between any two atoms, set at $4~\mu m$.
\item Geometric constraints defining the quantum register size (limited to a diameter of $100~\mu m$), ensuring that atom positions remain within the permissible physical area.
\end{itemize}

Each VC instance in our experiments required retraining the \textit{DEN} model from scratch. Despite this necessity, our heuristic proved significantly faster than exact optimization approaches. Specifically, the training procedure involved 3000 epochs per graph, executed with a batch size of 1, resulting in computational runtimes ranging between 1–2 minutes per graph for instances having $n \in [10, 16]$ vertices.

We also explore the robustness of our VC solution methods by considering cases where embeddings did not perfectly satisfy the UD criteria. To achieve this, we generated embeddings that fully complied with hardware constraints but might deviate from the ideal UD representation. This process resulted in test instances that retained the original graph's vertex count but had slightly altered connectivity (either missing or extra edges). Successfully solving VC problems on these imperfect embeddings provided empirical evidence demonstrating the practical robustness and adaptability of our quantum-based solution algorithms.

Once a feasible geometric embedding is obtained, the vertex positions are used to identify atoms and solve the resulting MWIS using a QAA algorithm over the embedded graph instances.

\subsubsection{Quantum Adiabatic-based Algorithm}

In our approach, we employed a QAA to solve the MWIS~\eqref{eq:objectif}--\eqref{eq:binaire}, following and extending the methods introduced in~\cite{aqmis, da2023quantum}.

We implemented the QAA by initializing the system in an easily prepared ground state, $|0\rangle^{\otimes n}$, and gradually evolving it toward the ground state of a final cost Hamiltonian $\mathcal{H}_C$. The evolution is governed by the adiabatic theorem, whereby the system remains close to its instantaneous ground state if the evolution is sufficiently slow~\cite{albash2018adiabatic}. Unlike strict adiabatic evolution, our heuristic allows for slight deviations into low-lying excited states, enabling a broader exploration of promising solution configurations.

To perform the evolution, we designed a pulse schedule that varies the detuning $\delta(t)$ and the Rabi frequency $\Omega(t)$ over time. Specifically, we set $\Omega(t_0) = 0$ and $\delta(t_0) < 0$ at the initial time $t_0$, and $\Omega(t_f) = 0$ and $\delta(t_f) > 0$ at the final time $t_f$. This results in an initial state of $|0\rangle^{\otimes n}$ and drives the system toward a low-energy subspace of $\mathcal{H}(t_f)$, which includes the ground state of $\mathcal{H}_C$.

To ensure that the quantum dynamics predominantly generate valid ISs, we estimated the interaction bounds of the system. Let $r_{ij}$ denote the distance between atoms $i$ and $j$ in the physical embedding of $\vertices$, and $C_6$ the device-specific interaction coefficient, we computed the following bounds:
\begin{align}
& R_{\min} = \underset{(i,j) \notin \edges}{\min}\quad r_{ij}, \
& r_{\max} = \underset{(i,j) \in \edges}{\max}\quad r_{ij}.
\end{align}

These bounds were used to define an admissible value for the Rabi frequency, $\Omega_b=C_6/r_b^6$ where $r_b=\sqrt{R_{\min}r_{\max}}$.
Given the hardware specifics, we also ensure that the value for $\Omega_b$ is within the admitted range $[0, \Omega_{\max}]$, $\Omega_{\max}=4\pi$, so $\Omega = \min(\Omega_b, \Omega_{\max})$.

\begin{figure}
    \centering
    \includegraphics[width=0.5\linewidth]{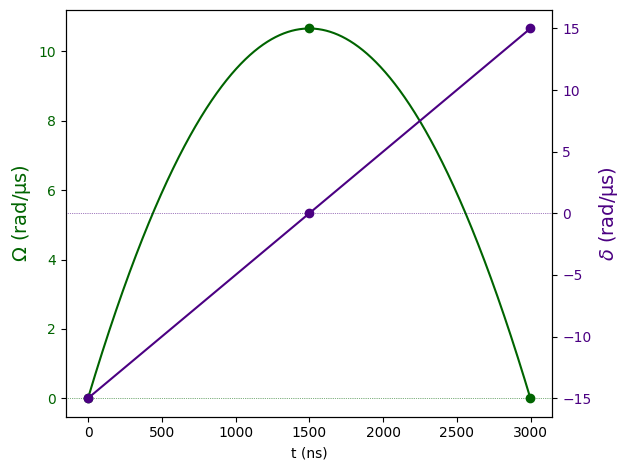}
    \caption{Adiabatic pulse example for a graph with $r_{\max} = 5.0\,\mu\text{m}$ and $R_{\min} = 8.7\,\mu\text{m}$.}
    \label{fig:pulse}
\end{figure}

Fig.~\ref{fig:pulse} shows an example of an adiabatic pulse for a graph instance with $r_{\max} = 5.0\,\mu\text{m}$ and $R_{\min} = 8.7\,\mu\text{m}$. The corresponding blockade radius, $r_b \approx 6.6\,\mu\text{m}$, sets the maximum value for the Rabi frequency to $\Omega_b = 10.66\,\text{rad}/\mu\text{s}$. During the HCG routine, the value of $\Omega_b$ is updated according to the current values of $r_{\max}$ and $R_{\min}$, which typically correspond to subgraphs of the original graph $\Graph$. The total evolution time is fixed at $3\,\mu\text{s}$, a duration compatible with the coherence time of the QPU. The detuning parameters are always initialized as $\delta(t_0) = -15\,\text{rad}/\mu\text{s}$ and ramped to $\delta(t_f) = +15\,\text{rad}/\mu\text{s}$, independently of the graph instance.

In the case of UD graphs, $\Omega $ allows preserving the independence property during the system's evolution~\cite{Pichler, quantumscars}. A higher value of $\Omega$ promotes broader exploration by accelerating transitions within the quantum state space. In the case of non-unit-disk graphs, this choice of $\Omega$ remains a practical trade-off. Specifically, if $ R_{\min} < r_{\max}$, the effective graph $\Graph'$ implicitly encoded by the quantum hardware contains more edges than the target graph $\Graph$, possibly excluding some feasible independent sets from the search space.
This property ensures that the pricing subproblem PSP remains reliable, as the quantum routine can still identify valid columns (i.e., independent sets) to improve the current RMP solution.

At each iteration of the column‐generation loop, the register design and the pulse shaping are called to solve the PSP; once no new columns can be generated, we hand the resulting solution off to the BB phase and invoke the primal heuristic to enforce integrality and obtain a complete coloring.

\subsection{Branch-and-Bound Method}
The Branch-and-Bound procedure systematically refines the solution of the VC problem by iteratively reducing the size of the graph fed to the RMP and fixing subsets of vertices to assigned colors. This is achieved through three key operations: \textit{branching}, \textit{bounding}, and \textit{pruning}.

\subsubsection{Branching}
At each node of the BB tree, we invoke the HCG routine to generate a collection of maximal independent sets.  From each mIS~$\tilde S$, we create a child subproblem by fixing $\lambda_{\tilde S}=1$ in the set–partitioning formulation~\eqref{eq:cg:obj}–\eqref{eq:cg:binary}.  Concretely, if the parent node corresponds to subgraph $\Graph'=(\vertices',\edges')$, branching on $\tilde S$ produces
\[
\Graph'' = \bigl(\,\vertices'\setminus\tilde S\,,\;\edges'\setminus\{(i,j)\mid i\in\tilde S\text{ or }j\in\tilde S\}\bigr),
\]
thereby assigning one color to all vertices in~$\tilde S$ and reducing the remaining coloring problem to~$\Graph''$.

\subsubsection{Bounding}
For each node, we compute a valid lower bound on its chromatic number by combining:
\begin{itemize}[nosep]
  \item The node depth $d$ (the number of colors already fixed),
  \item The continuous relaxation bound from the RMP $LB_{RMP}$,
  \item Classical spectral bounds on~$\chi(\Graph'')$, i.e., \ Hoffman’s bound $LB_{H}$~\cite{hoffman}, the Elphick–Wocjan bound $LB_{EW}$~\cite{elphick2016inertial}, and the Edwards–Elphick bound $LB_{EE}$~\cite{edwards1983lower}.
\end{itemize}
We then set

\begin{equation}
\mathrm{LB}_{\text{node}} = d + \max\left\{ LB_{\text{RMP}},\; LB_{\text{H}},\; LB_{\text{EW}},\; LB_{\text{EE}} \right\}
\end{equation}

where we take the maximum among the available lower bounds, since we are solving a minimization problem and aim to obtain the tightest possible bound.

\subsubsection{Pruning}\label{sec:pruning}

To explore only the most relevant regions of the solution space, we apply pruning rules that allow us to discard nodes unlikely to lead to improved solutions.

\begin{description}[nosep]
  \item[Non–Improving Pruning (Rule~1):] Discard any node whose $\mathrm{LB}_{\text{node}}$ exceeds or equals the best known feasible coloring, i.e., the $\mathrm{UB}$.
  \item[Redundancy Pruning (Rule~2):] If the same subgraph $\Graph''$ has already been processed in another branch, discard this duplicate node.
\end{description}

\subsubsection{Node Selection.}
We prioritize exploration by assigning each active node a score
\[
\mathrm{score}(\Graph'') \;=\; \bigl(\text{UB}_{\text{loc}}(\Graph'')\bigr)\;\times\;\bigl|\edges''\bigr|,
\]
where $\text{UB}_{\text{loc}}(\Graph'')$ is the feasible coloring obtained via our greedy primal heuristic. Nodes with higher scores are explored first, guiding the search efficiently toward optimality.

\subsection{Primal Heuristic}

Within the BP framework, a primal heuristic is generally employed at each node to rapidly derive a feasible integer solution from the set of available columns, i.e., ISs generated during the column generation step. Given that global optimality is enforced via the BB procedure, an exact integer resolution of the RMP is unnecessary at each node. Therefore, we introduce a fast greedy heuristic that constructs a feasible coloring solution from the existing columns.

The primal heuristic proceeds as follows:
\begin{enumerate}[label=(\roman*),nosep]
    \item Sort the vertices in the current subgraph $\Graph'$ by decreasing degree;
    \item Select the highest-degree vertex $v$;\label{step:heur}
    \item From the available ISs, choose the largest set containing vertex $v$, and assign it as a new color class;
    \item Remove all vertices included in this chosen IS from further consideration;
    \item Repeat steps~\ref{step:heur}–(iv) until all vertices in $\Graph'$ are colored.
\end{enumerate}

This straightforward yet sufficient heuristic provides a valid coloring, yielding a reliable local upper bound ($\mathrm{UB}_{\text{loc}}$) on the chromatic number at each node. By leveraging exclusively the ISs generated through column generation, it avoids additional computational overhead from supplementary quantum (QPU) or classical (ILP) solver calls, thus ensuring efficient integration within the overall QCBP algorithm.

\subsection{The Overall Hybrid QCBP Algorithm}

\begin{figure}[ht]
    \centering
    \includegraphics[width=0.7\linewidth]{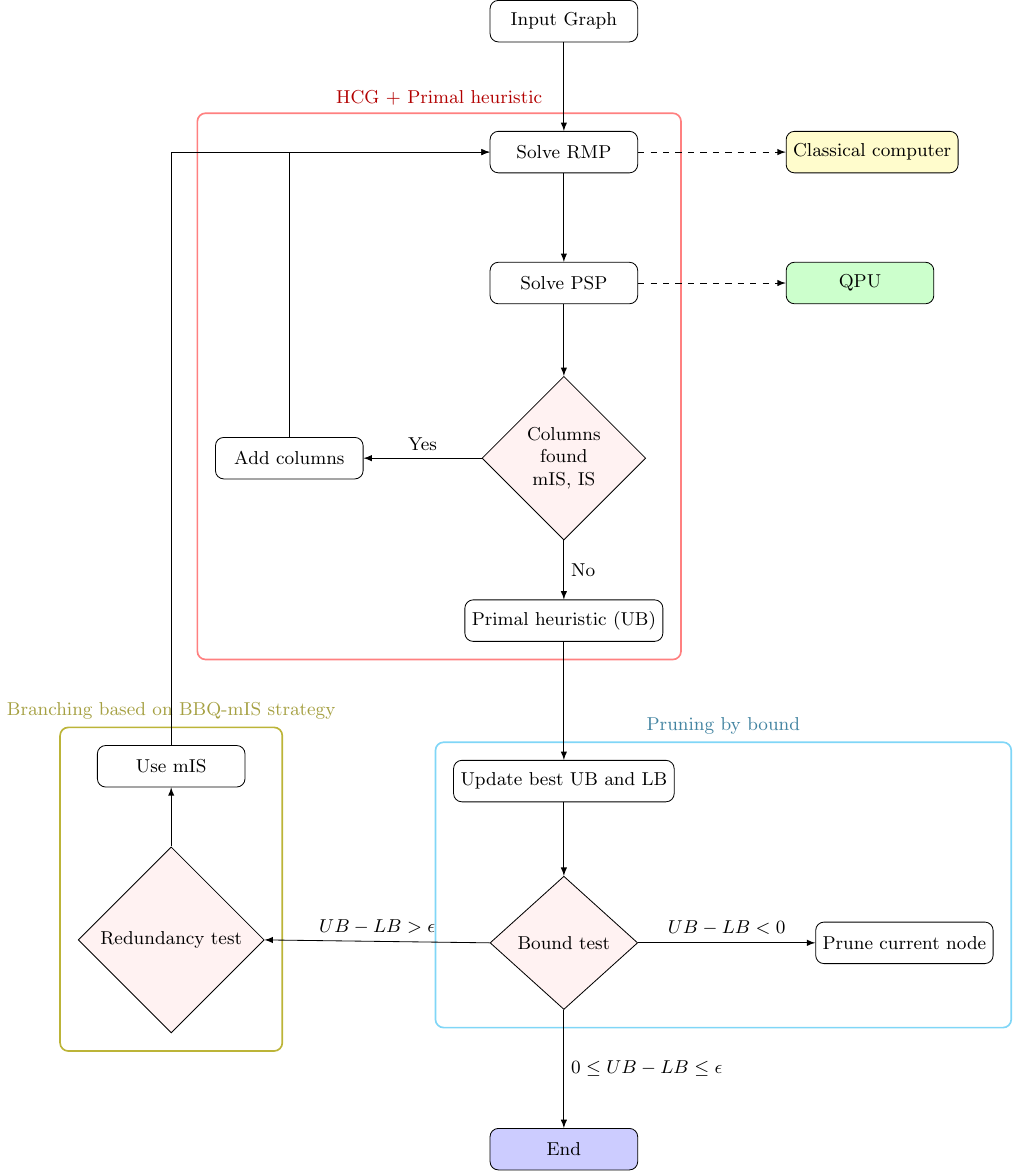}
    \caption{Hybrid Quantum-Classical Branch-and-Price Algorithm (QCBP).}
    \label{fig:hybrid-Branch-and-Price}
\end{figure}

A schematic overview of the complete QCBP algorithm is depicted in Fig.~\ref{fig:hybrid-Branch-and-Price}. 
The procedure begins by solving the initial Restricted Master Problem defined over the original graph $\Graph$, and proceeds iteratively through the Hybrid Column Generation phase, combining quantum sampling and classical exact methods to effectively generate promising columns. Upon termination of the column generation process, the set of identified independent sets serves as input for the primal heuristic and initiates the branching phase of the Branch-and-Bound procedure. Each branch of the resulting BB tree corresponds to assigning a specific color to a selected maximal independent set, progressively narrowing the feasible solution space toward integer optimality.

The QCBP algorithm leverages the complementary strengths of quantum and classical computational approaches, striking a favorable balance between solution optimality and quantum resource utilization. As it will turn out in the next section, it achieves substantially improved solution quality compared to HCG while significantly reducing the quantum computational overhead relative to conventional hybrid methods such as BBQ-mIS alone. 

\subsection{Advantages of QCBP over Existing Approaches}

The QCBP algorithm is developed by combining and leveraging effective elements from both BBQ-mIS and HCG methods to address their individual limitations. The primary objectives are to enhance the probability of reaching optimal solutions and reduce the number of QPU calls, a critical factor due to the practical constraints of quantum hardware execution.

In quantum emulation settings, computational costs increase significantly with the number of qubits because of the numerical complexity involved in solving the Schr\"odinger equation. Conversely, these costs remain relatively constant with respect to the number of measurement shots, as these shots are drawn from a fixed, precomputed probability distribution. On real QPU hardware, the computation time is mainly determined by the hardware’s sampling rate (approximately 2 Hz) making it largely independent of qubit count. Consequently, minimizing QPU calls directly reduces total computational time.

\subsubsection{Limitations of BBQ-mIS}
The BBQ-mIS approach requires extensive QPU resources due to repeated QAOA executions involving pulse-level parameter optimization. Specifically, previous implementations permitted up to 50 optimization function evaluations, each needing 200 QPU shots to estimate the objective function accurately, plus an additional 200 shots for the final evaluation. Hence, each BB node potentially required up to:
\[
200 \times 50 + 200 = 10{,}200 \text{ QPU samples},
\]
a substantial computational cost when scaled across multiple nodes. Moreover, BBQ-mIS employs general lower bounds that typically underestimate the actual chromatic number, leading to an increased exploration of the BB tree and thus further sampling requirements.

\subsubsection{Limitations of HCG}
The primary limitation of the HCG method lies in its relatively lower solution accuracy, notably when addressing non-UD graphs. This limitation arises mainly due to the absence of a further classical optimization step, which could otherwise refine and improve solution quality. However, by employing the QAA rather than QAOA, HCG considerably reduces quantum resource demands by avoiding the necessity of circuit parameter optimization.

\subsubsection{QCBP: Combining Strengths}
QCBP addresses the above limitations by integrating effective aspects from both prior approaches. Firstly, it adopts tighter lower bounds derived from the relaxed RMP solutions employed by HCG. These bounds have demonstrated superiority over those used in BBQ-mIS, effectively reducing the number of BB nodes evaluated.

Secondly, QCBP eliminates QAOA from the branching process. Instead, it directly utilizes mISs generated during the PSP solution within HCG iterations. These mISs directly produce child nodes in the BB tree without additional quantum optimization calls.

Each iteration of the HCG in QCBP involves:
\begin{enumerate}[label=\roman*)]
    \item Solving the relaxed RMP,
    \item Solving the MWIS PSP via QAA,
    \item If QAA yields no valid independent set, using classical ILP via the GLPK solver.
\end{enumerate}
The iterative process stops when no additional columns (independent sets) can be found, either through quantum or classical methods.

Upon completion of the column generation phase, a primal heuristic calculates a feasible coloring solution, providing an upper bound on the chromatic number. Concurrently, the relaxed RMP solution delivers a valid lower bound. Branching is then performed exclusively on maximal independent sets. Each branch fixes the corresponding vertices in an mIS, thus creating subproblems represented by reduced graphs.

The feasibility of solutions generated through branching is ensured by construction, negating the need for additional validation checks. Nodes are pruned based on the previously defined pruning rules utilizing both the RMP-based and classical theoretical bounds. The BB procedure continues recursively, invoking HCG at each node until one of the following stopping conditions is met:
\begin{itemize}[nosep]
    \item The predefined maximum number of BB nodes is reached,
    \item No further maximal independent sets remain available for branching,
    \item Optimality is proven when the upper bound equals the lower bound.
\end{itemize}

With the complete algorithmic framework specified, we next evaluate QCBP’s solution quality and quantum resource efficiency through rigorous experimental comparisons.

\section{Results}
\label{sec:results}

In this section, we evaluate the performance of the QCBP algorithm using classical emulation and real quantum hardware (the Orion Alpha neutral-atom QPU). We analyze three key metrics: optimality rates, quantum resource efficiency (shot counts), and scaling behavior.

\subsection{Benchmark Instances and Experimental Setup}\label{sec:dataset}

We constructed a benchmark dataset of 140 VC instances to evaluate QCBP under noiseless quantum conditions, based on \textit{classical emulation} of the QAA for solving the PSP subproblem. The dataset comprises 20 graphs for each value of number of vertices $n \in \{10, 11, 12, 13, 14, 15, 16\}$. Among these, 78 graphs have a UD representation fully-compliant with hardware constraints, while the remaining 62 graphs are non-UD, exhibiting slight deviations from ideal representations. We deliberately included these non-perfect graphs to assess algorithmic robustness, which is crucial in realistic NISQ environments.

In addition to the emulation dataset, we constructed a second benchmark specifically designed for testing on real \textit{quantum hardware}. This smaller dataset was used to evaluate performance on the Orion Alpha QPU. It consists of UD graphs with increasing size: 3 graphs with $n = 10$ vertices, 1 with $n = 20$, 2 with $n = 30$, and 1 with $n = 40$ vertices. Unlike the smaller emulation instances, the graphs with more than 10 vertices could not be efficiently simulated classically due to their size and computational complexity. Using this dataset, we assessed the quality of the QPU-generated solutions by analyzing whether the sampled configurations represented good candidate solutions to the VC problem.

\subsection{Results from Quantum Emulation}

\subsubsection{Optimality Performance}

We first assess the optimality rates of QCBP compared to BBQ-mIS and HCG, using the exact ILP solver GLPK as reference for the VC problem solution. Table~\ref{tab:optimality} summarizes the success rates. As we can see, QCBP outperforms both alternative methods, failing in only 3 out of 140 instances. Moreover, QCBP shows much greater robustness to non-UD graph representations, where performance drops severely for other methods — from about 97\% to 73\% in BBQ-mIS, and down to only 40\% in HCG..

\begin{table}[ht!]
\centering
\caption{Percentage of instances solved optimally by each method.}
\begin{tabular}{|c|c|c|c|}
\hline
 Unit-Disk & BBQ-mIS & HCG & QCBP\\
 \hline
True & 0.923 & 0.897 & 0.987 \\
False & 0.726 & 0.403 & 0.968 \\
\hline
Total & 0.836 & 0.679 & 0.979 \\
\hline
\end{tabular}
\label{tab:optimality}
\end{table}

Table~\ref{tab:error} reports the average optimality gaps, computed as $(c - \chi(\Graph))/\chi(\Graph)$, where $c$ denotes the solution obtained. QCBP consistently achieves near-optimal solutions (gap = 0 in most cases), requiring at most one additional color beyond the optimum. Its worst-case gap is 0.033, compared to 0.208 for HCG and 0.083 for BBQ-mIS.

\begin{table}[ht!]
\centering
\caption{Average relative optimality gap grouped by graph size $n$.}
\begin{tabular}{|c|c|c|c|}
\hline
 & BBQ-mIS & HCG & QCBP\\
 \hline
$n=10$ & 0.033 & 0.050 & 0.0 \\
$n=11$ & 0.050 & 0.067 & 0.0 \\
$n=12$ & 0.083 & 0.117 & 0.0 \\
$n=13$ & 0.017 & 0.108 & 0.033 \\
$n=14$ & 0.075 & 0.092 & 0.0 \\
$n=15$ & 0.050 & 0.150 & 0.0 \\
$n=16$ & 0.083 & 0.208 & 0.017 \\
\hline
\end{tabular}
\label{tab:error}
\end{table}

\subsubsection{Quantum Resource Efficiency}

\begin{figure}[!htb]
  \centering
  \begin{subfigure}[b]{0.44\textwidth}
    \centering
    \includegraphics[width=\linewidth]{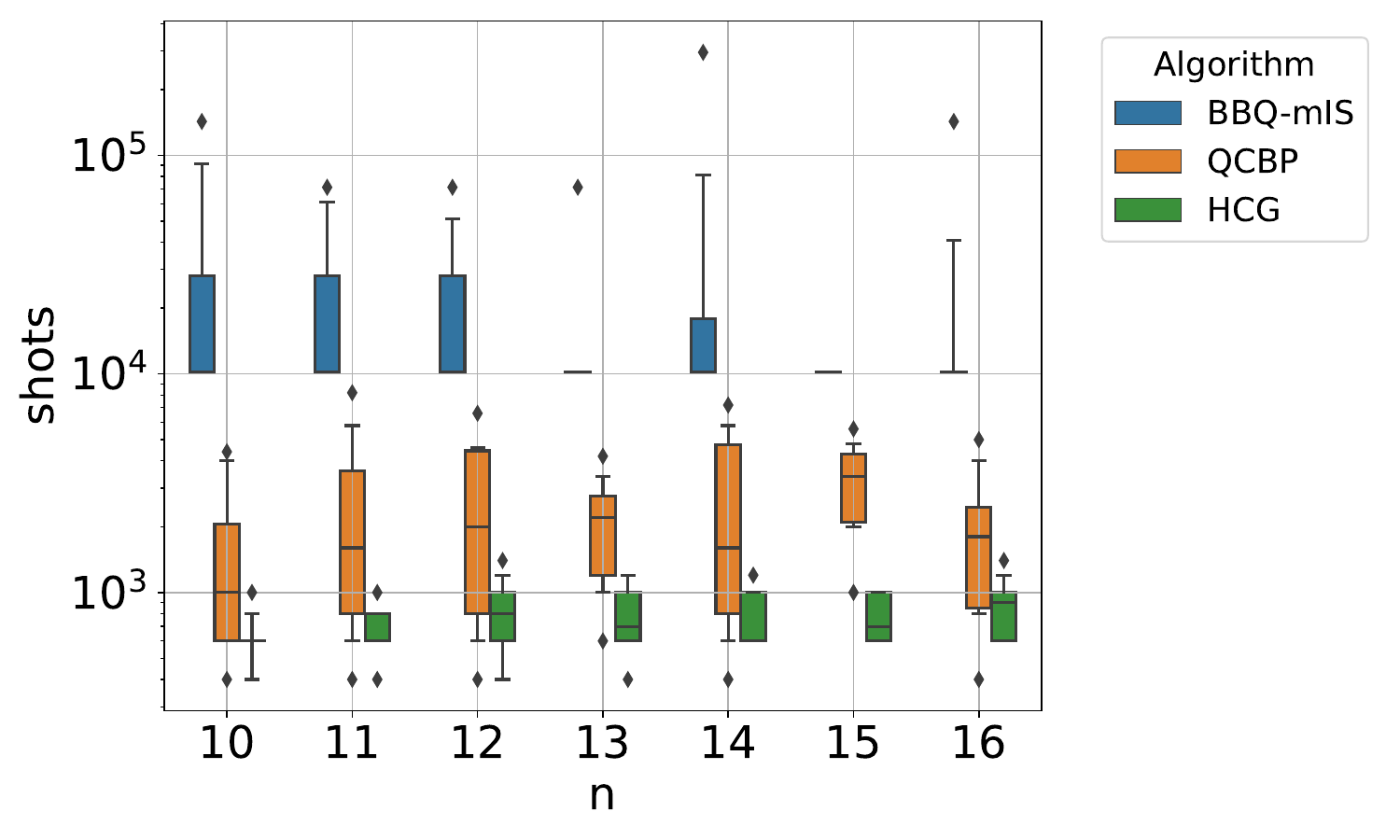}
    \caption{UD graph instances.}
    \label{fig:QPU_UD}
  \end{subfigure}\hfill
  \begin{subfigure}[b]{0.44\textwidth}
    \centering
    \includegraphics[width=\linewidth]{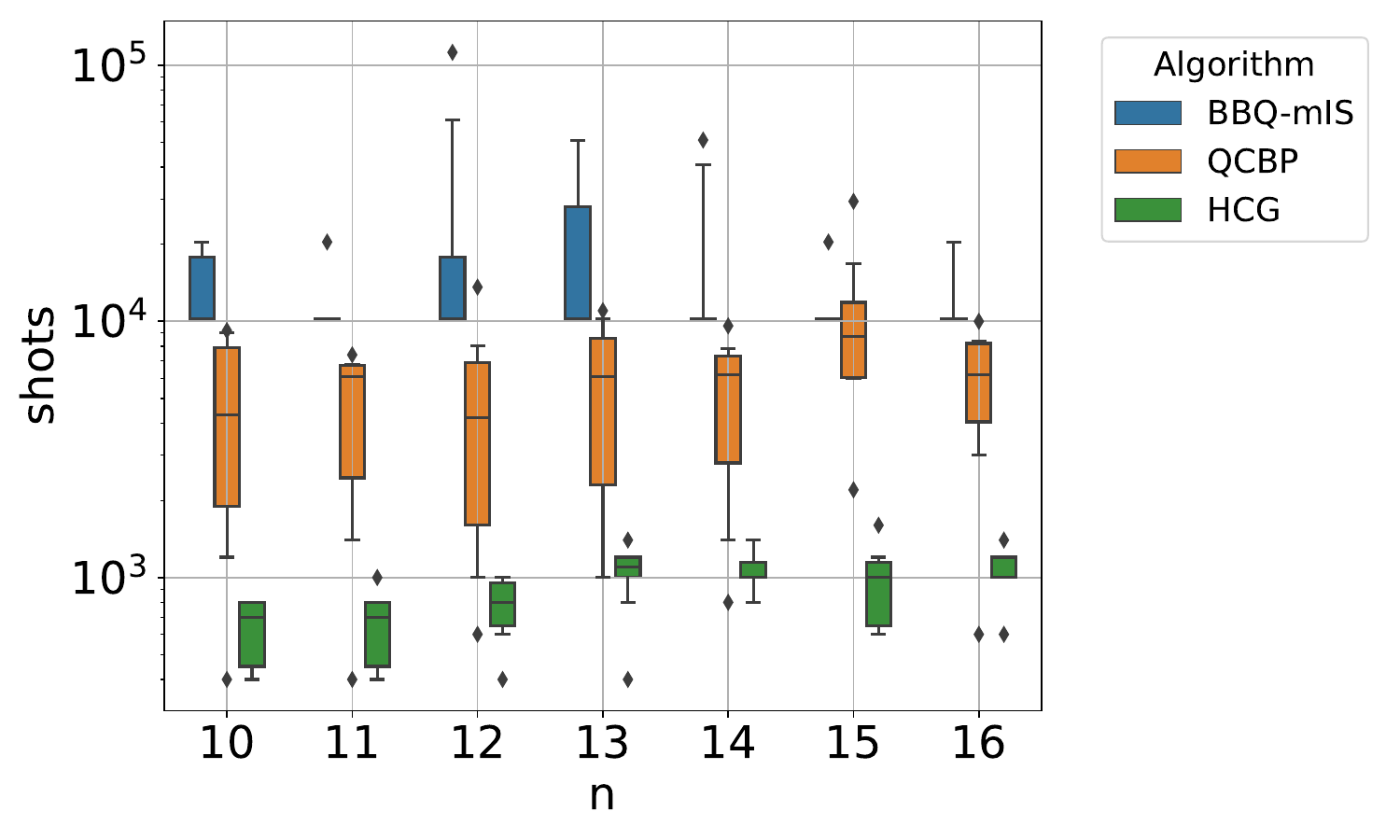}
    \caption{Non-UD graph instances.}
    \label{fig:QPU_noUD}
  \end{subfigure}
  \caption{Quantum shot distributions by method and graph type.}
  \label{fig:QPU}
\end{figure}

Beyond solution quality, it is also important to consider the number of QPU calls, i.e., the quantum samples or shots required to solve each VC instance. Fig.~\ref{fig:QPU} compares shot requirements across methods, grouped by $n$ and separated into UD (Fig.~\ref{fig:QPU_UD}) and non-UD (Fig.~\ref{fig:QPU_noUD}) graphs. HCG requires the fewest shots ($10^2$--$10^3$), though at the cost of lower solution quality. BBQ-mIS, while effective, consumes significantly more shots ($10^4$--$10^5$), often an order of magnitude higher than HCG. QCBP, in contrast, achieves a favorable balance, maintaining high accuracy with far fewer QPU shots. This is particularly evident in UD graphs (Fig.~\ref{fig:QPU_UD}), where QCBP's shot distribution ($<10^4$) is much closer to HCG than to BBQ-mIS. Such efficiency is crucial because quantum resources are limited, so achieving high-quality solutions with fewer shots directly improves the practicality and scalability of solving VC problems on current quantum hardware.

\subsubsection{Classical Optimization Analysis}


The two main classical components of the QCBP algorithm to analyze are the Branch-and-Bound nodes generated during the search and the calls to the ILP exact solver to complement the quantum solution of the PSP. 
Examining the distribution of BB nodes provides insight into the quality of the bounds computed during the search, while the number of ILP calls reflects the effectiveness of the quantum method within the HCG routine: fewer ILP calls indicate that the MWIS is being solved efficiently.

\begin{figure}[!htb]
  \centering
  \begin{subfigure}{0.32\textwidth}
    \includegraphics[width=\linewidth]{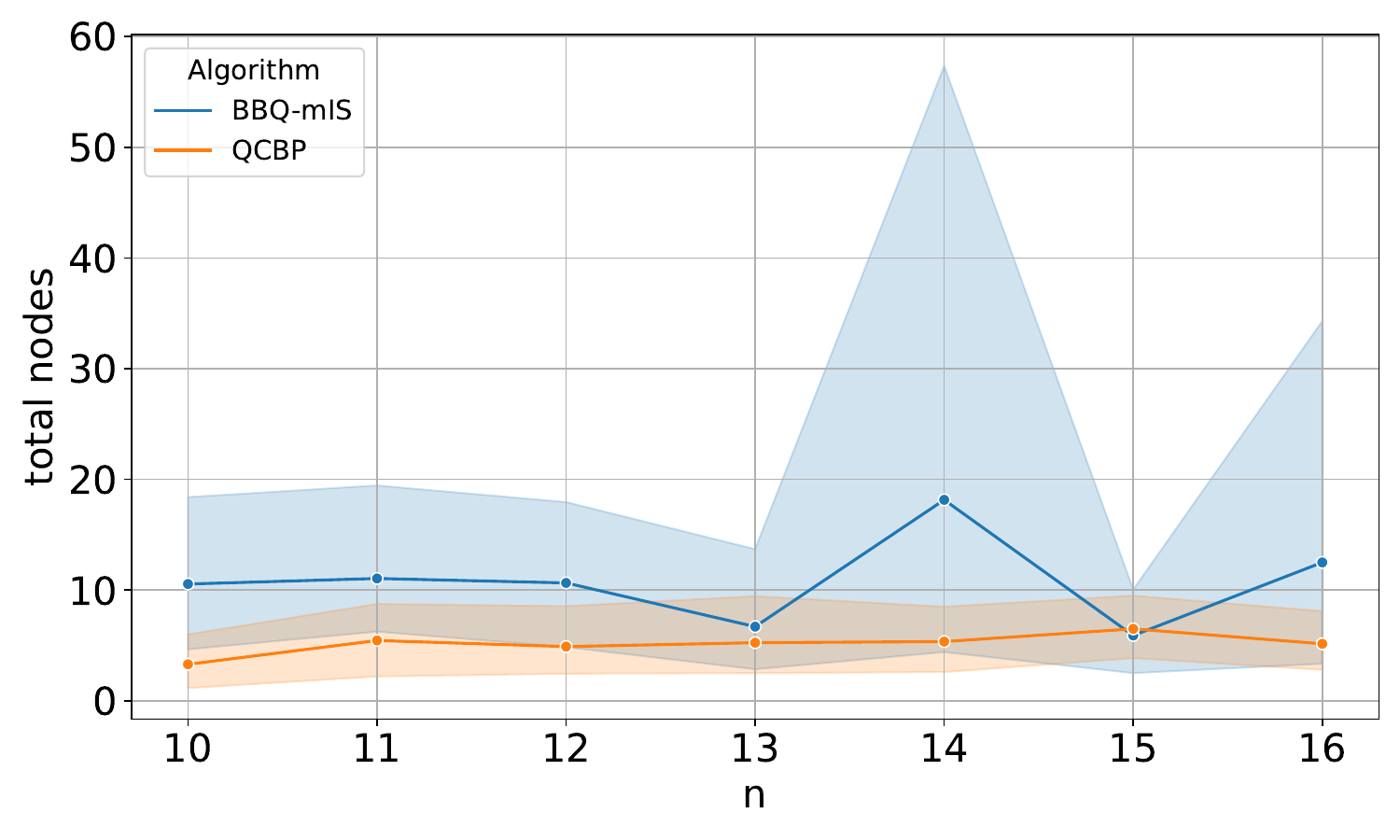}
    \caption{Total generated BB nodes.}
    \label{fig:nodes_total}
  \end{subfigure}\hfill
  \begin{subfigure}{0.32\textwidth}
    \includegraphics[width=\linewidth]{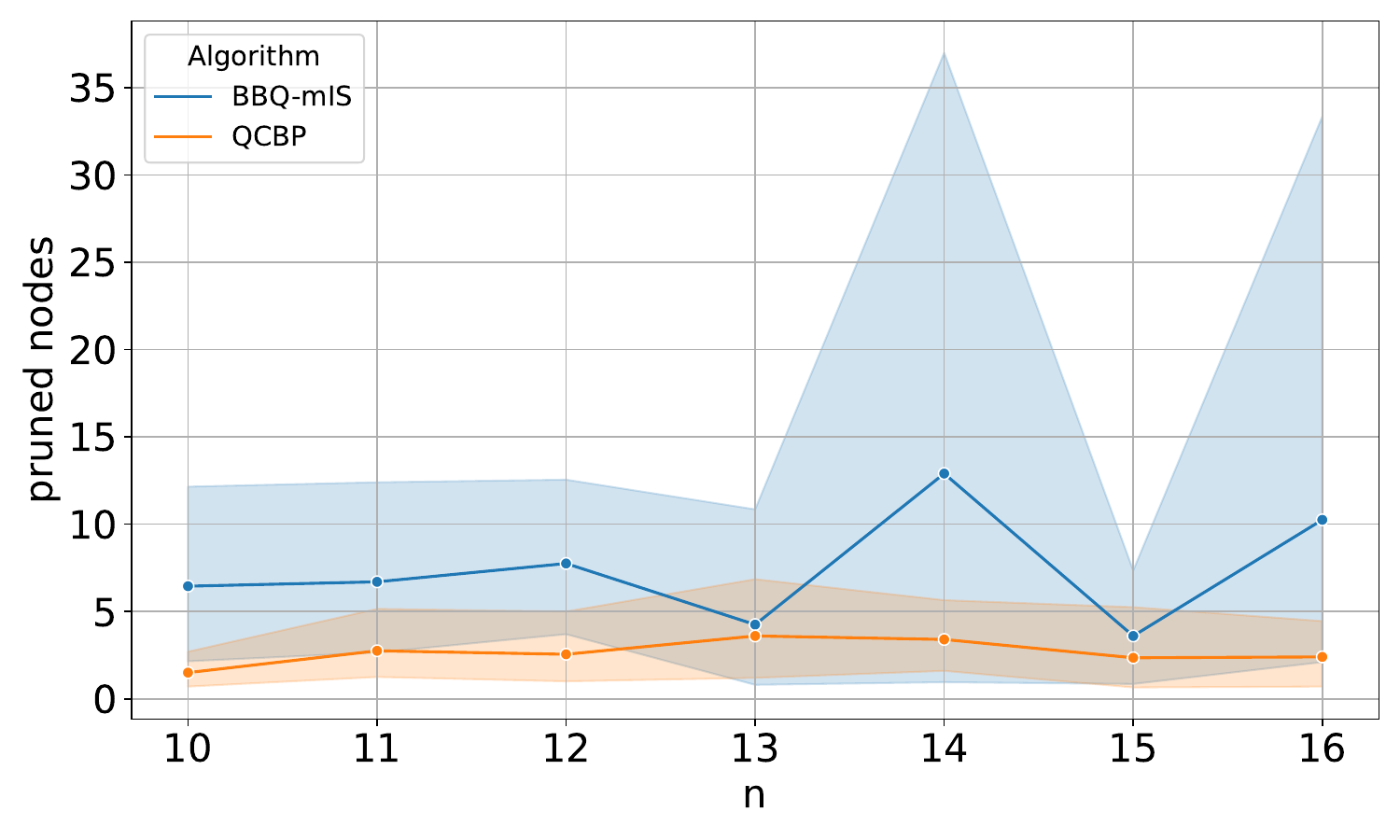}
    \caption{Pruned BB nodes.}
    \label{fig:nodes_pruned}
  \end{subfigure}\hfill
  \begin{subfigure}{0.32\textwidth}
    \includegraphics[width=\linewidth]{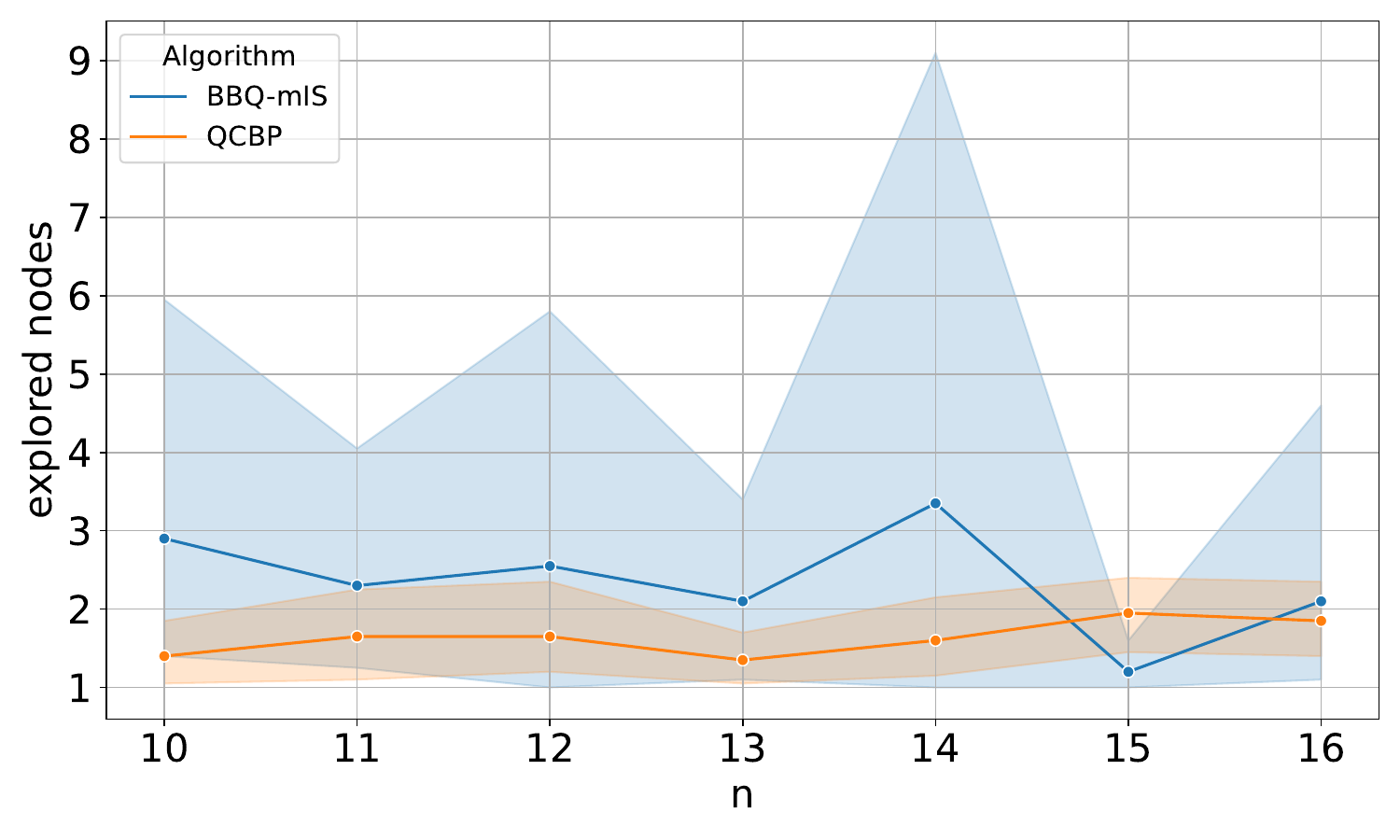}
    \caption{Explored BB nodes.}
    \label{fig:nodes_explored}
  \end{subfigure}
  \caption{Branch-and-Bound node distributions grouped by graph size $n$.}
  \label{fig:nodes}
\end{figure}

Fig.~\ref{fig:nodes} presents detailed statistics on BB node activity. For each value of $n$, we report the range and median number of BB nodes generated, explored, and pruned (see Section~\ref{sec:pruning} and Fig.~\ref{fig:nodes_pruned}). BBQ-mIS consistently generates and explores far more nodes—a node is considered \textit{explored} when the HCG is invoked to compute a new objective—due to its weaker lower bounds. This leads to greater variability in node counts, with total nodes reaching up to 60 (compared to 10 for QCBP) and explored nodes up to 9 (versus 3 for QCBP). In contrast, QCBP leverages tighter bounds from relaxed RMPs, enabling more aggressive pruning and earlier termination. As a result, it maintains a nearly constant number of explored nodes across instances and solves problems more efficiently.
\captionsetup{aboveskip=2pt, belowskip=0pt}
\captionsetup[sub]{aboveskip=2pt}

Concerning the number of calls to the ILP exact solver within the HCG routine for QCBP, minimizing these calls is important to reduce classical computational overhead. Table~\ref{tab:ILP} reports the median number of ILP calls for each value of $n$, separately for UD and non-UD graphs. ILP calls are more frequent for non-UD instances, where the physical qubit connectivity does not perfectly represent the edge structure of the induced graph, highlighting a limitation of relying exclusively on quantum solvers such as QAA for the PSP step. ILP usage generally increases with problem size—except for $n=16$—ranging from 1 to 7 in the UD case and from 11 to 21.5 in the non-UD case. Future work could explore incorporating classical heuristics to solve the MWIS PSP more efficiently, either independently or as a post-processing step to enhance QAA-generated solutions.

\begin{table*}[ht!]
\centering
\caption{Median number of exact ILP solver calls in QCBP.}
\begin{tabular}{|c|c|c|c|c|c|c|c|}
\hline
Unit-Disk & $n=10$ & $n=11$ & $n=12$ & $n=13$ & $n=14$ & $n=15$ & $n=16$\\
\hline
True & 1.0 & 4.0 & 3.5 & 4.0 & 4.5 & 7.0 & 3.0\\
False & 11.0 & 13.5 & 10.0 & 13.0 & 14.5 & 21.5 & 12.5 \\
\hline
\end{tabular}
\label{tab:ILP}
\end{table*}

\subsection{Evaluation on real neutral-atom QPU}

Given the promising emulation results, we tested larger graphs on the Orion Alpha neutral-atom QPU, using the UD instances described in \ref{sec:dataset}.
The aim of this study is to evaluate whether QCBP is robust to quantum noise and capable of efficiently addressing the MWIS problem as the PSP within the HCG procedure, by identifying sets with negative reduced cost. In addition, we assess whether the identified sets are maximal, which is important for the branching component. Another key aspect is to demonstrate that the entire algorithm can operate in a closed-loop manner, efficiently alternating between QPU calls through direct access and the classical computational steps.

\begin{figure}[ht!]
    \centering
    \includegraphics[width=0.4\linewidth]{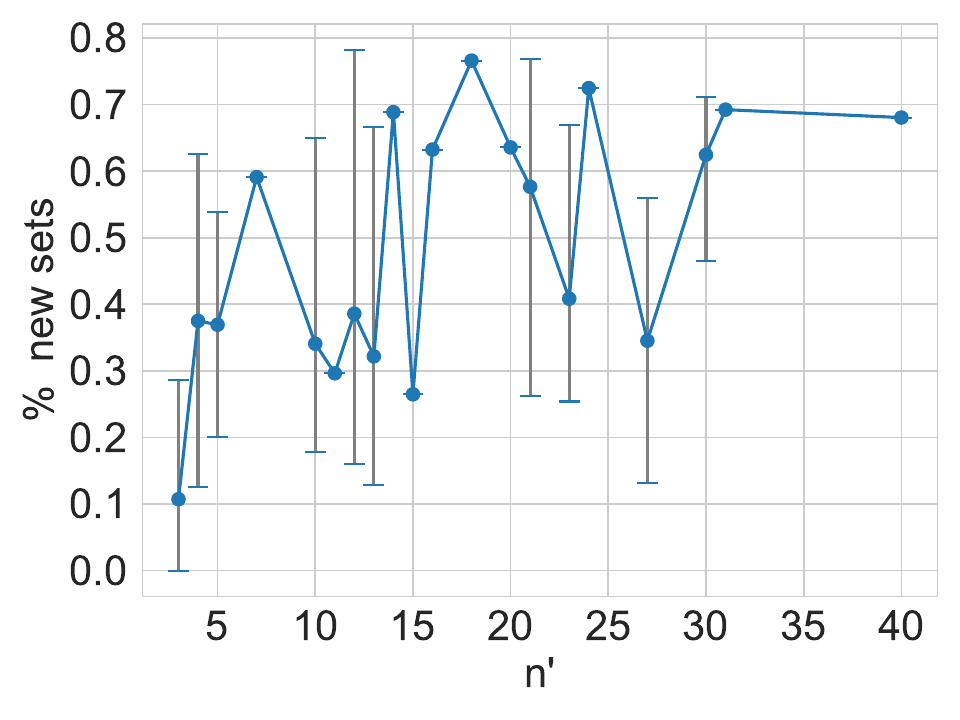}
    \caption{Percentage of new sets found through PSP solution via QPU, grouped by PSP size $n'$.}
    \label{fig:perc_new_sets}
\end{figure}

Fig.~\ref{fig:perc_new_sets} illustrates the percentage of relevant PSP solutions: those with negative reduced cost \(1 - \sum_{i \in \vertices} \pi^*_i z_i < 0\), identified via the QPU relative to the total number of samples collected. The results are grouped according to the size \(n'\) of the MWIS subproblem solved within the HCG framework. Although the original graph dimension is \(n\), the subproblem size \(n'\) can be smaller due to exclusion of vertices with negative dual values or branching in the BB process.
As shown, a high percentage of new PSP solutions are obtained, particularly for larger subproblems (\(n' > 15\)), with up to nearly 80\% of the bitstrings corresponding to valid PSP solutions. This demonstrates the effective sampling capabilities of QCBP on real hardware, especially for unit-disk graphs, where MWIS sampling can efficiently identify large independent sets. Conversely, the proportion of new solutions decreases for smaller subproblems (\(n' \leq 5\)), falling below 40\%, which may be attributed to the limited availability of additional valid sets or their prior inclusion in earlier iterations.



\begin{figure}[ht!]
    \centering
    \includegraphics[width=0.4\linewidth]{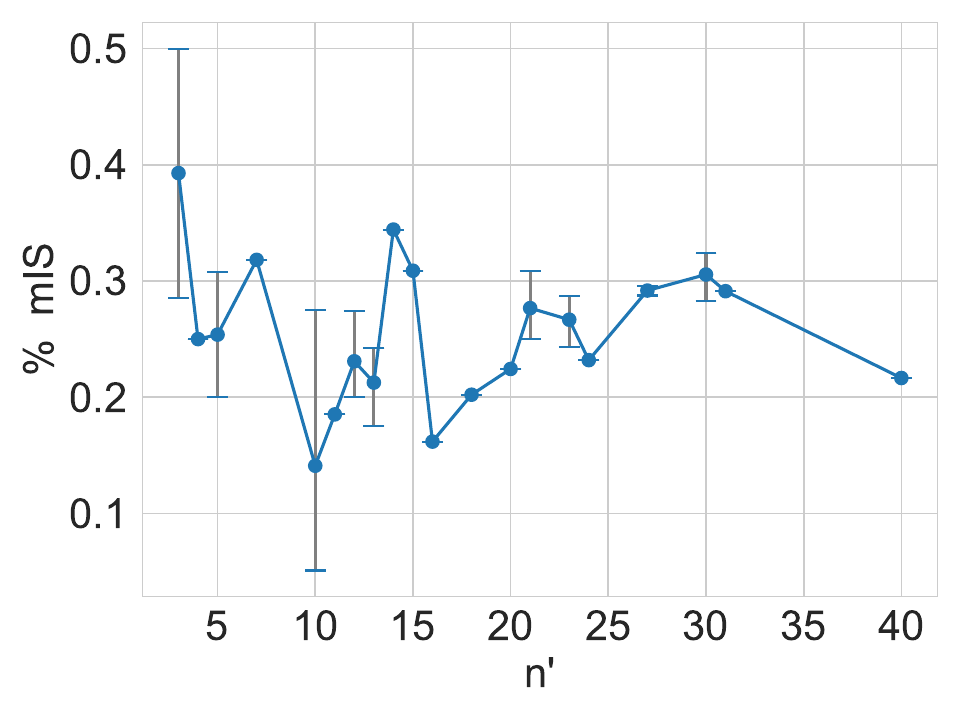}
    \caption{Percentage of mISs found through PSP solutions via QPU, grouped by PSP size $n'$.}
    \label{fig:perc_mIS}
\end{figure}

Concerning the branching procedure, generating branches based on mISs is particularly important. To evaluate the QCBP algorithm’s effectiveness in identifying such sets, we analyze the percentage of mISs found as a function of the subproblem dimension $n'$ within the PSP. Fig.~\ref{fig:perc_mIS} illustrates these results, showing that while the percentage of identified mISs decreases as $n'$ increases, due to the growing difficulty of including all relevant vertices, the proportion remains substantial.

\begin{figure}[!htb]
  \centering
  \begin{subfigure}[b]{0.48\textwidth}
    \centering
    \includegraphics[width=0.9\linewidth]{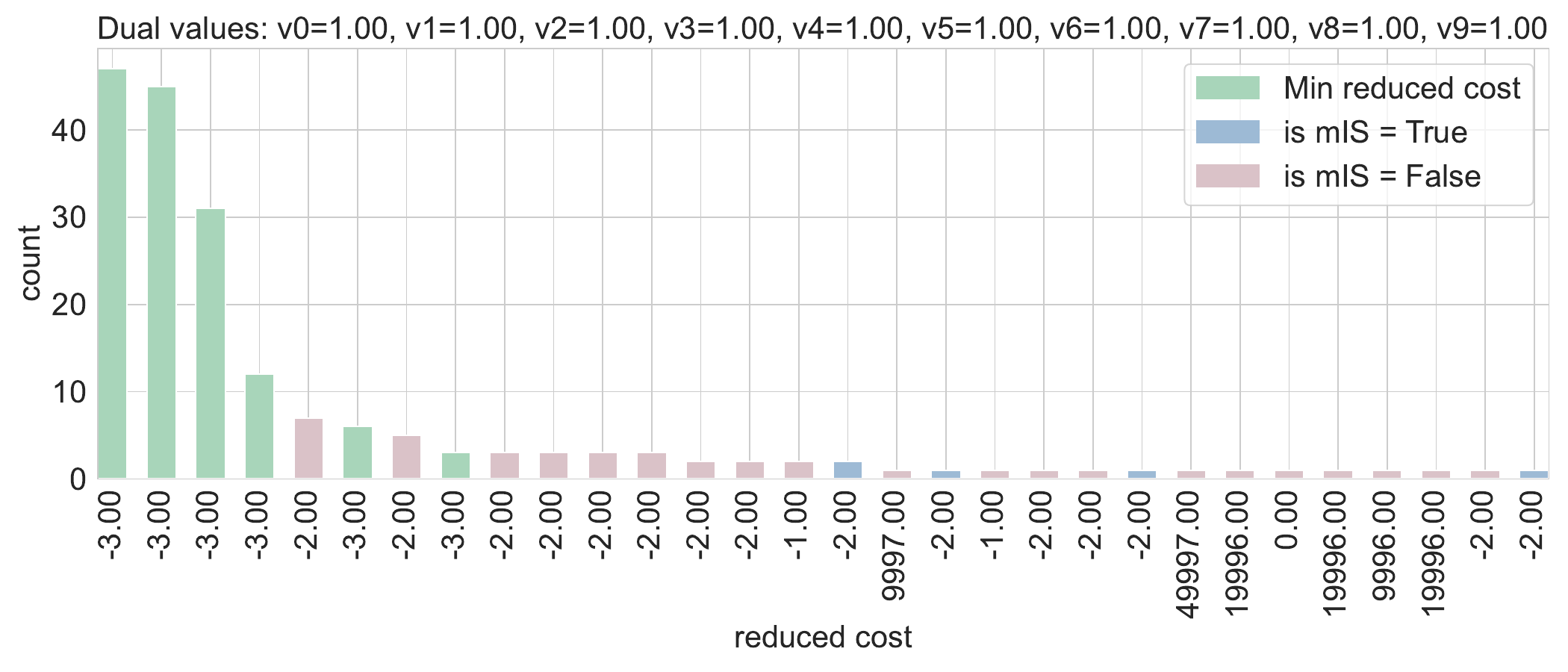}
    \caption{First HCG iteration: PSP solved with Orion Alpha QPU.}
    \label{fig:hcg1}
  \end{subfigure}\hfill
  \begin{subfigure}[b]{0.48\textwidth}
    \centering
    \includegraphics[width=0.9\linewidth]{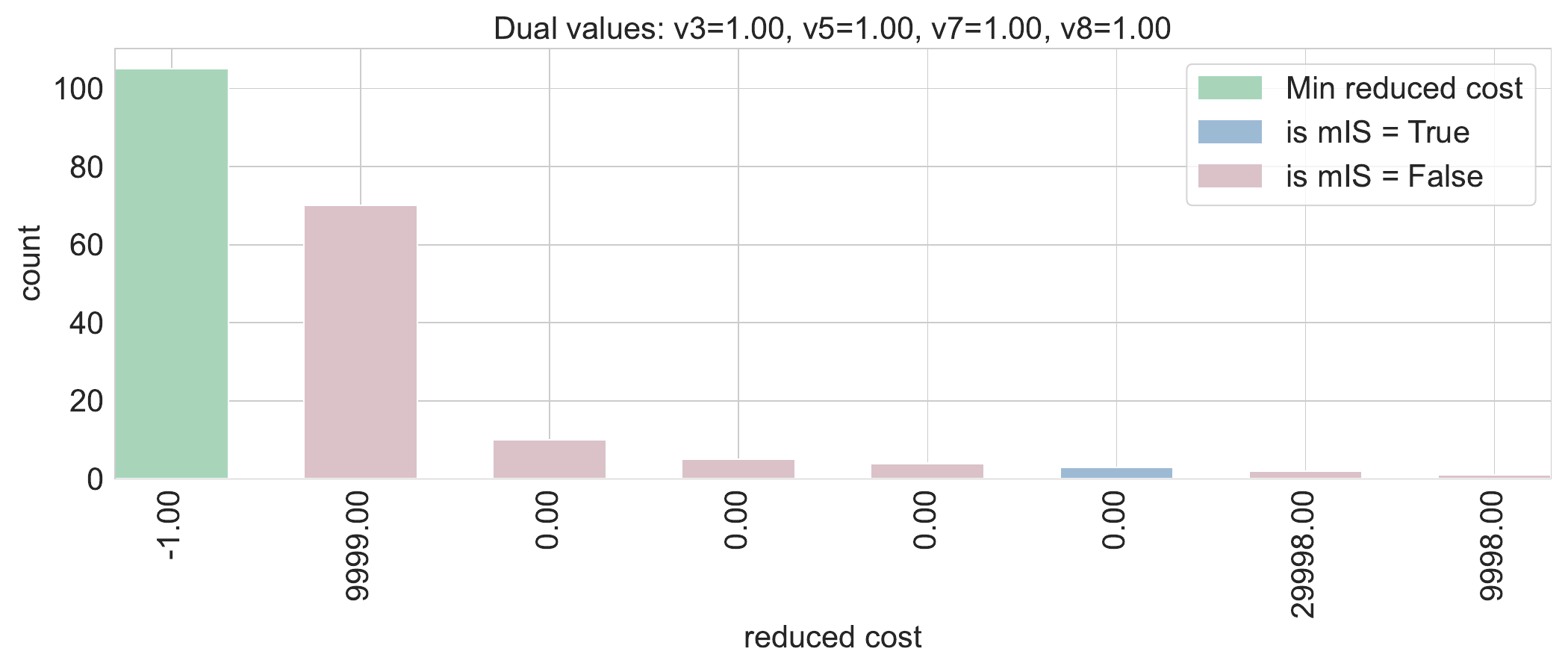}
    \caption{Second HCG iteration: PSP solved with Orion Alpha QPU.}
    \label{fig:hcg2}
  \end{subfigure}

  \medskip 

  \begin{subfigure}[b]{0.48\textwidth}
    \centering
    \includegraphics[width=0.9\linewidth]{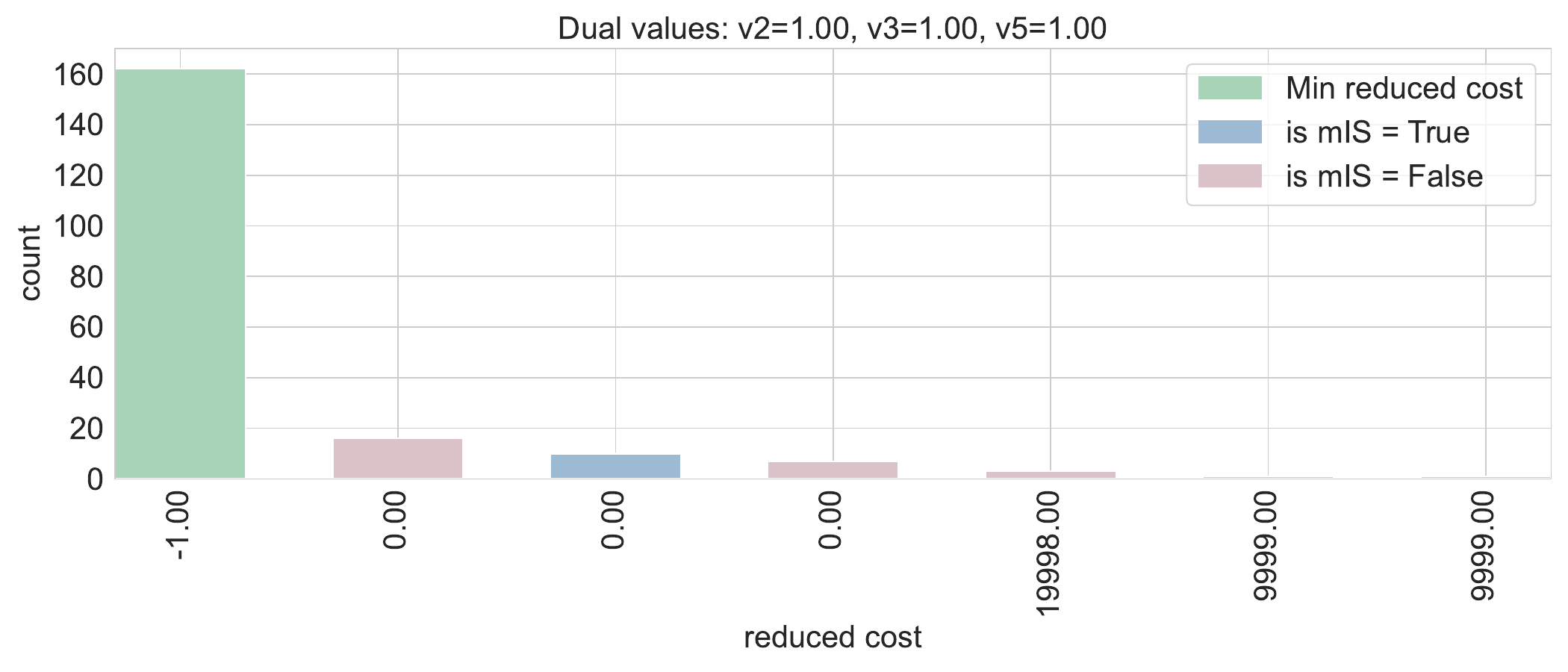}
    \caption{Third HCG iteration: PSP solved with Orion Alpha QPU.}
    \label{fig:hcg3}
  \end{subfigure}\hfill
  \begin{subfigure}[b]{0.48\textwidth}
    \centering
    \includegraphics[width=0.9\linewidth]{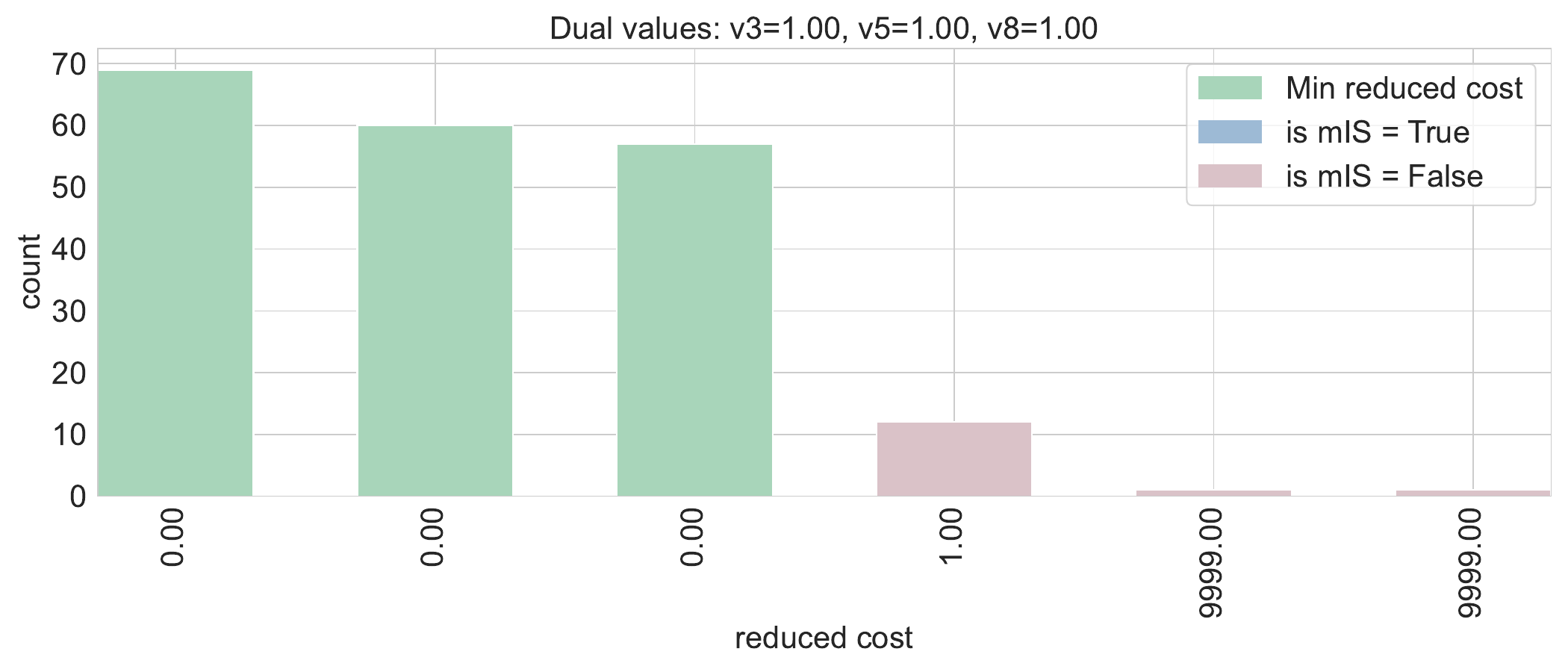}
    \caption{Fourth HCG iteration: PSP solved with Orion Alpha QPU.}
    \label{fig:hcg4}
  \end{subfigure}

  \caption{Occurrences of each bitstring sampled by the QPU and corresponding reduced cost.}
  \label{fig:qpu_example}
\end{figure}

\begin{figure}[ht!]
    \centering
    \includegraphics[width=0.5\linewidth]{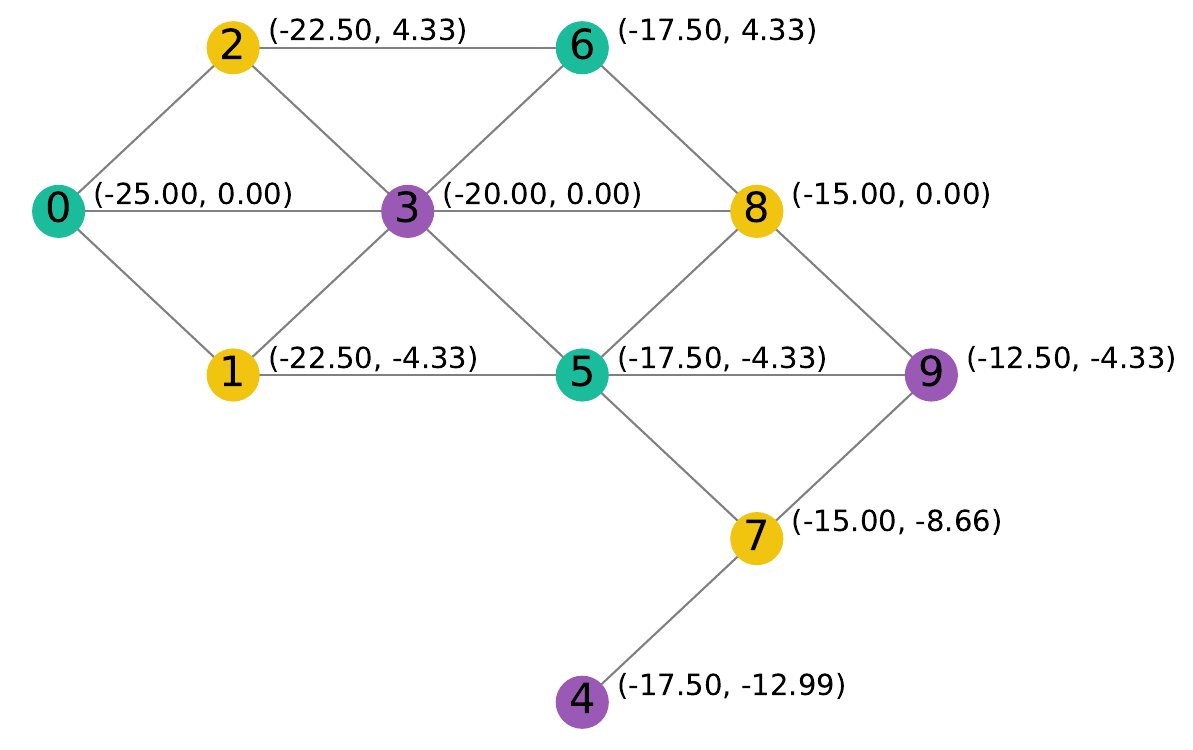}
    \caption{Final coloring of a 10-vertex VC instance solved with QCBP on Orion Alpha QPU.}
    \label{fig:qpu_coloring}
\end{figure}

A full example of the QCBP algorithm applied to a 10-vertex instance is detailed in Figures~\ref{fig:qpu_example}--\ref{fig:qpu_coloring}, illustrating the workflow: iterative column generation via QPU sampling, termination verification, heuristic-based coloring, and final optimality confirmation. Specifically, the HCG routine was executed for 4 iterations, each aimed at identifying new sets with negative reduced cost to enrich subsequent RMPs. 
In the initial iteration (Fig.~\ref{fig:hcg1}), the QPU sampler using the QAA algorithm performed 200 samples over the full graph \(\Graph\), yielding predominantly negative reduced-cost samples, with 24 new sets identified out of 30. Subsequent iterations (Figs.~\ref{fig:hcg2}--\ref{fig:hcg4}) operated on progressively smaller subgraphs, defined by the positive dual values, with the QPU identifying new valid columns until no further improvements were found, i.e. when all the samples have reduced costs $\geq0$ (see Fig. \ref{fig:hcg4}), as confirmed by a final call to the exact ILP solver.

\captionsetup{font=small}
\captionsetup[sub]{font=small}

Finally, the primal heuristic combined all discovered columns to produce a feasible coloring using 3 colors, matching the size of the largest clique in \(\Graph\) and confirming the optimal solution.

This result demonstrates that the complete QCBP workflow, from quantum sampling to classical post-processing and optimization, successfully identifies optimal solutions on practical instances.

\section{Conclusion}
\label{sec:conclusion}

In this paper, we introduced the Quantum Classical Branch-and-Price method, a novel hybrid algorithm designed to efficiently solve the Vertex Coloring problem on neutral-atom quantum processors. Our computational results demonstrate that QCBP significantly outperforms existing approaches, such as BBQ-mIS and HCG, achieving optimal solutions in approximately 98\% of benchmark instances. While QCBP slightly increases quantum resource usage (measured in QPU calls) compared to the HCG method, it remains substantially more efficient than BBQ-mIS. The observed computational efficiency arises primarily from leveraging the Quantum Approximate Algorithm to sample high-quality maximal independent sets and from reducing the number of Branch-and-Bound nodes explored before reaching termination. We validated the practical viability of QCBP through experiments on real QPUs, where the algorithm reliably identified optimal solutions despite quantum noise and hardware limitations. These findings underscore the robustness and scalability of QCBP for addressing larger and more complex graph instances.

Future research avenues include improving solution optimality by tightening lower bounds within the Branch-and-Bound framework and enhancing the quality of mIS solutions obtained from quantum samplers. A promising direction is the classical post-processing of non-maximal independent sets into valid maximal solutions through targeted vertex inclusion or exclusion. Finally, the methodological framework introduced here is not limited to Vertex Coloring and can be generalized to tackle various other combinatorial optimization problems through hybrid quantum-classical approaches.

\bibliographystyle{IEEEtran}
\bibliography{biblio}

\end{document}